\documentclass[10pt,aps,pra,twocolumn,superscriptaddress,floatfix,nofootinbib]{revtex4-1}
\makeatletter
\def\@bibdataout@aps{%
 \immediate\write\@bibdataout{%
  @CONTROL{%
   apsrev41Control,author="08",editor="1",pages="0",title="0",year="1",eprint="1"%
  }%
 }%
 \if@filesw
  \immediate\write\@auxout{\string\citation{apsrev41Control}}%
 \fi
}%
\makeatother 
\usepackage{lipsum}
\usepackage{amsmath}
\usepackage{amsfonts, amssymb, bbm, braket, accents}
\usepackage{graphicx}   
\usepackage[usenames,dvipsnames]{color}
\usepackage[makeroom]{cancel}
\usepackage{soul} 

\newcommand{\ain}{a_{{\rm in}}}
\newcommand{\aout}{a_{{\rm out}}}

\newcommand{\opin}[1]{#1_\textrm{in}}
\newcommand{\opout}[1]{#1_\textrm{out}}

\newcommand{\conj}[1]{{#1}^{*}}

\newcommand{\dg}{^\dagger}

\newcommand{\barG}[1]{\bar{\Gamma}_{#1}}
\newcommand{\conG}[1]{\Gamma_{#1}^{*}}
\newcommand{\Gam}[1]{\Gamma_{#1}}

\newcommand{\expt}[1]{\langle {#1} \rangle}

\newcommand{\op}[2]{\left |{#1}\right\rangle\! \!\left \langle {#2}\right |}
\newcommand{\defeq}{\mathrel{:=}}

\usepackage[breaklinks=true]{hyperref}
\hypersetup{
  colorlinks   = true, 
  urlcolor     = blue, 
  linkcolor    = blue, 
  citecolor   = red 
}
\usepackage{amsthm}
\usepackage{cleveref} 
\crefformat{equation}{Eq.~(#2#1#3)} 
\crefformat{section}{Sec.~#2#1#3} 
\Crefformat{equation}{Equation~(#2#1#3)}
\crefformat{figure}{Fig.~#2#1#3}
\crefrangeformat{equation}{Eqs.~#3(#1)#4--#5(#2)#6}

\begin{document}

\title{Two-photon self-Kerr nonlinearities for quantum computing and quantum optics}

\author{Joshua Combes}
\email{joshua.combes@gmail.com}
\affiliation{Centre for Engineered Quantum Systems, School of Mathematics and Physics, University of Queensland, Brisbane, QLD, Australia}

\author{Daniel J. Brod}
\email{danieljostbrod@id.uff.br}
\affiliation{Instituto de F\'isica, Universidade Federal Fluminense, Av. Gal. Milton Tavares de Souza s/n, \\
Gragoat\'a, Niter\'oi, RJ, 24210-340, Brazil}

\date{\today}

\begin{abstract}
The self-Kerr interaction is an optical nonlinearity that produces a phase shift proportional to the square of the number of photons in the field.
At present, many proposals use nonlinearities to generate photon-photon interactions. 
For propagating fields these interactions result in undesirable features such as spectral correlation between the  photons.
Here, we engineer a discrete network composed of cross-Kerr interaction regions to simulate a self-Kerr medium. The medium has effective long-range interactions implemented in a physically local way.
We compute the one- and two-photon S matrices for fields propagating in this medium. 
From these scattering matrices we show that our proposal leads to a high fidelity photon-photon gate. 
In the limit where the number of nodes in the network tends to infinity, the medium approximates a perfect self-Kerr interaction in the one- and two-photon regime.
\end{abstract}

\maketitle

\section{Introduction}\label{sec:intro}
The self-Kerr effect is a photon-number-dependent nonlinearity. In a single bosonic mode, where $[a,a\dg]=1$, the unitary corresponding to a self-Kerr interaction with strength $\chi$ and duration $t$ is  $U(t)= \exp(i\chi t a\dg a\dg a a)$. At the two-photon level, the transformation induced in the Fock basis is
\begin{align}
\alpha \ket{0}+\beta \ket{1} +\gamma \ket{2} \rightarrow \alpha \ket{0}+\beta \ket{1} +\gamma e^{i2\chi t}\ket{2},
\end{align}
where $\phi = 2\chi t$ is the phase shift acquired by the two-photon component. There are many interesting applications of such an interaction, e.g.\ the creation of cat states~\cite{StobMilbWodk08}, parameter estimation \cite{GenoInvePari09,RossAlbaPari16}, quantum devices~\cite{Roy10}, construction of qubits~\cite{Mabu12}, and quantum gates~\cite{PachChou00,KnilLaflMilb01,NystMcCuHeuc17}. 

Independent of these applications, there is a long and rich theoretical history of studying the field theory of scattered photons from a spatially-extended self-Kerr medium~\cite{DrumCart87,LaiHaus89,Wrig90,ChiaDeutGarr91,BlowLoudPhoe91,DeutChiaGarr93,JoneShap93,AbraCohe94,BoivKartHaus94}. Due to experimental advances \cite{WeisKungDumu15,Park16}, there is renewed interest in such a medium in the context of Rydberg vapors~\cite{GorsOtteFlei11,BienBuch16,LahaFirs17}, coupled nonlinear cavities~\cite{LeeNohSche15,SeeNohAnge17,PedePlet17}, and even scattering from point-like Kerr interactions~\cite{WaksVuck06,Kosh08,LiaoLaw10,XuRephFan13,NystMcCuHeuc17}.

In this multimode (field-theoretic) setting, some applications could have limitations at the few photon level, as pointed out by~\citet{Shap06} and \citet{GeaBana10}. One central aspect of these objections could be formalized by the cluster decomposition principle (CPD) \cite{XuRephFan13,XuFan17,SancCadaMart18}. Consider the scattering of (at most) two photons off a nonlinear and spatially-localized system. In general the single-photon scattering matrix is
\begin{align} \label{eq:singleSdelta}
S_{\omega,\nu} = t_{\omega} \delta(\omega-\nu),
\end{align}
indicating a frequency dependent phase shift $t_\omega$ when input $\nu$ and output $\omega$ frequencies are equal. The two-photon S-matrix connecting input frequencies $\nu_1$ and $\nu_2$ to output frequencies $\omega_1$ and $\omega_2$ can be decomposed in two terms
\begin{align} \label{eq:twoSdelta}
S_{\omega_1,\omega_2,\nu_1,\nu_2}  = S^0_{\omega_1,\omega_2,\nu_1,\nu_2} + i T_{\omega_1,\omega_2,\nu_1,\nu_2}.
\end{align}
The first term represents an energy-conserving process for non-interacting photons,
\begin{align}\label{eq:twoSdeltaS}
S^0_{\omega_1,\omega_2,\nu_1,\nu_2} =& t_{\omega_1} t_{\omega_2}[ \delta(\omega_1-\nu_1)\delta(\omega_2-\nu_2)\nonumber\\
&+ \delta(\omega_1-\nu_2)\delta(\omega_2-\nu_1)],
\end{align}
where $t_{\omega_k}$ is the same as in \cref{eq:singleSdelta}, and the symmetrized delta functions are due to bosonic statistics. The second term arises from photon-photon interactions mediated by the systems in the scattering region
\begin{align} \label{eq:twoSdeltaT}
T_{\omega_1,\omega_2,\nu_1,\nu_2} =& C_{\omega_1,\omega_2,\nu_1,\nu_2} \delta(\omega_1+\omega_2-\nu_1-\nu_2),
\end{align}
 where $C_{\omega_1,\omega_2,\nu_1,\nu_2}$ is a function of the denoted frequencies as well as various system parameters. The frequency constraint imposed by the delta function encodes energy conservation. A corollary of the CDP is that the $T$-matrix (and therefore the $S$-matrix) cannot~\cite{XuRephFan13,XuFan17}, for a local scattering site, be of the form
\begin{align} \label{eq:idealSmatrix}
C_{\nu_1,\nu_2}[ \delta(\omega_1-\nu_1)\delta(\omega_2-\nu_2) + \delta(\omega_1-\nu_2)\delta(\omega_2-\nu_1)].
\end{align}
However, for some applications the desired S-matrix is exactly of the above form. Specifically in optical quantum computing, the ideal S-matrix for a controlled-phase gate would be \cref{eq:idealSmatrix} with $C_{\nu_1,\nu_2} = e^{i\phi}$, where $\phi$ is the phase shift (see Eq.\ (15) of \cite{XuRephFan13}). Thus, it has been argued that it is impossible to construct a photon-photon interaction in a way that would lead to a high fidelity momentum-based phase gate \cite{XuRephFan13}.

In this paper, we describe a physical setup that circumvents the restrictions imposed by the cluster decomposition principle. 
To that end we employ two tricks: (i) we use a spatially-distributed medium, and (ii) we mimic counter-propagation in one chiral mode.  As the length of the medium goes to infinity we formally break the assumptions of the CDP. The effective counter-propagation allows us to avoid the spectral entanglement, which can be traced back to momentum conservation \cite{BrodComb16a}.

\begin{figure}[ht]
\includegraphics[width=\columnwidth]{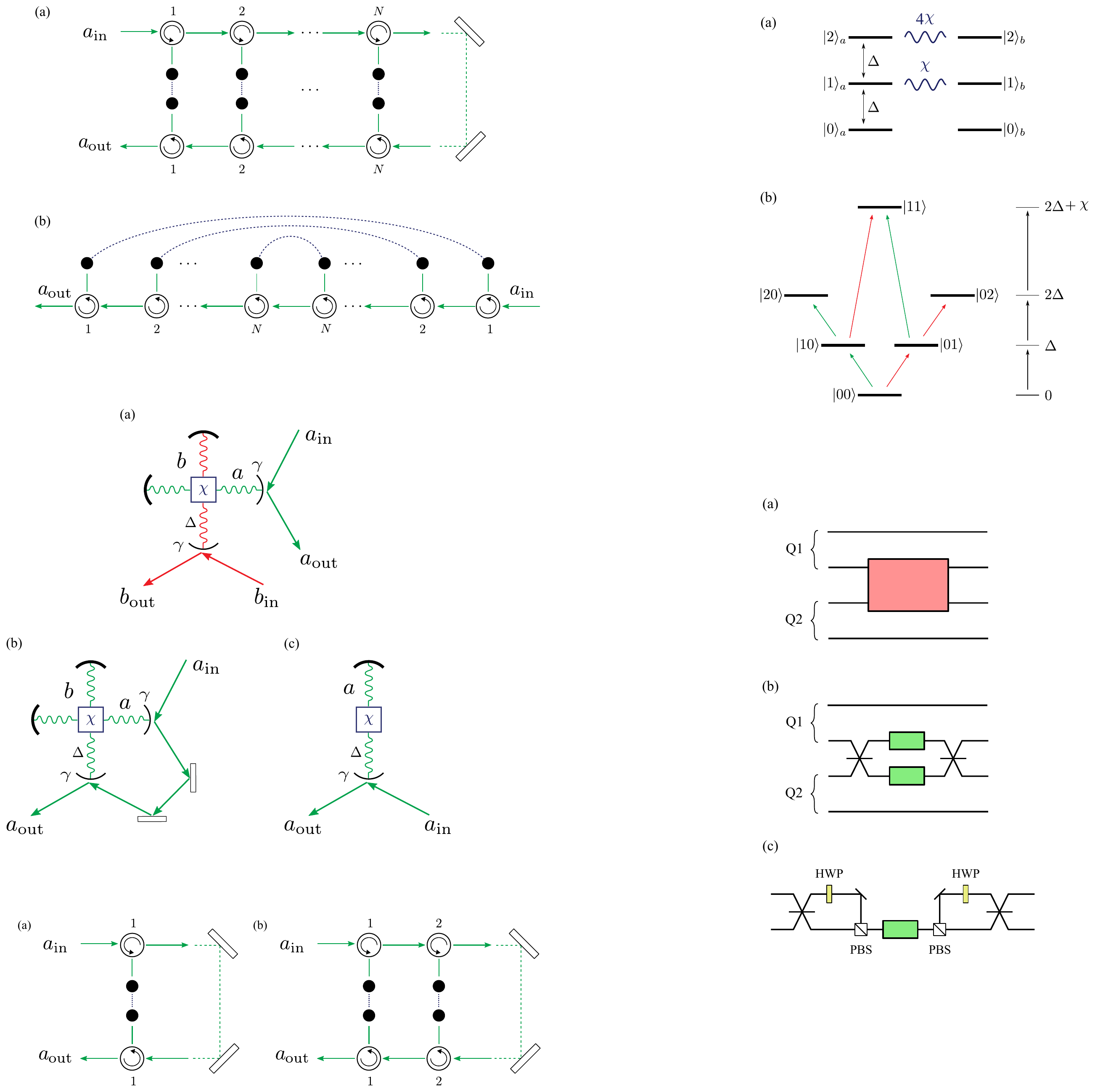}
\caption{(Color Online). (a) An interaction between two input fields is mediated by a cavity containing a cross-Kerr medium, described by Hamiltonian $H_{\rm int} =  \chi a\dg a  b\dg b$. The cavity decay rate is $\gamma$ and the cavity Hamiltonian is $\Delta a\dg a$. (b) By feeding one output mode of the cavity into the other input mode, we expect to obtain an effective self-Kerr medium. (c) A cavity with a self-Kerr medium, such that $H_{\rm int} = \chi  a\dg a a\dg a$, which is equivalent to $H_{\rm int} =\chi  a\dg  a\dg a a$ up to a term proportional to the number operator.}\label{fig:cavities}
\end{figure}

\begin{figure}[ht]
\includegraphics[width=\columnwidth]{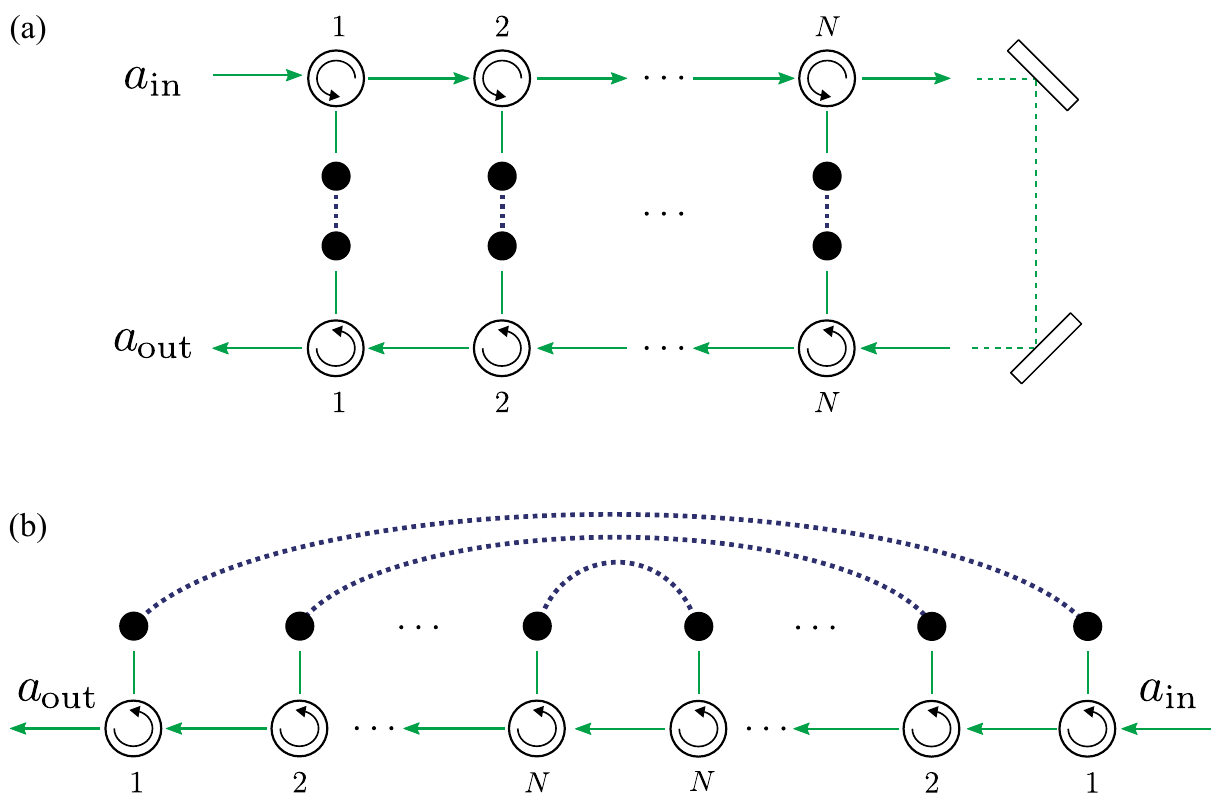}
\caption{(Color Online). (a)  A chain of $N$ interaction sites. The unit cell in \cref{fig:cavities}(a) is represented differently here. Black dots represent cavity modes, while the purple dashed lines represent the cross-Kerr interaction. We enforce chirality of the fields by using circulators, which we represent by circles with arrows. By cascading $N$ interaction sites 
and connecting the output of the upper chain into the input of the lower chain, we simulate counter-propagating conditions, the basis of the proposal in \cite{BrodComb16a,BrodComb16b}. (b) The cascading in (a) can also be seen effectively as a local implementation of a series of nonlocal interactions, similar to a spin ladder-type interaction \cite{GopaRiceSigr94,Soly00,Mila00,ZhouKanoNg17}.}\label{fig:nsites}
\end{figure}

Unfortunately, an infinitely long medium is unphysical. Thus we consider a system composed of a 1D chain of $N$ interaction sites. Each interaction site has a cavity containing a cross-Kerr medium, see \cref{fig:cavities} and Refs.~\cite{BrodComb16b,BrodComb16a}, and supports two modes that couple independently to left- and right-propagating fields. At the end of the 1D chain there is a mirror which feeds back the output of the last interaction site into itself, see \cref{fig:nsites} (a). This effectively gives rise to a one-input, one-output system. The mirror mimics a counter-propagating arrangement by bouncing the chiral field back into the chain, propagating in the opposite direction. Moreover the counter-propagation can be interpreted as turning physically-local interactions into effectively nonlocal ones, as represented in \cref{fig:nsites} (b). When combined, our approaches to points (i) and (ii) allow us to engineer a self-Kerr nonlinearity for at most two propagating photons with a fidelity (relative to the ideal process) that increases in the number of interaction sites.

Finally we use our multimode self-Kerr medium to construct a nonlinear sign shift gate~\cite{KnilLaflMilb01}. We were motivated by the recent proposal of \citet{NystMcCuHeuc17} which achieved a fidelity of $F=0.84$ with two two-level emitters. Here we show we show that with three sites (which could be constructed from a total of six 3-level atoms) we get $F=0.95$ and generally we can approach $F=1$ as the number of sites increases.

\subsection{Notation and conventions}\label{sec:notation}

In this paper, we consider an external field which couples to local physical systems in different ways, such as exemplified in \cref{fig:cavities}. We denote the input and output field operators by $\ain$ and $\aout$, respectively. The field couples to cavity operators, which we denote by $a$ and $b$ satisfying $[a,a\dg]=1=[b,b\dg]$ and $[a,b\dg]=0$, with some coupling strength $\sqrt{\gamma}$. When we consider more than one physical interaction site, we denote the operators of site $i$ as $a_i$ and $b_i$, satisfying the obvious commutation relations. The cavities in the interaction sites have resonance frequencies $\Delta$, and contain cross-Kerr Hamiltonians with interaction strength $\chi$. 

\section{Theoretical Background: Input-Output theory, cascading and SLH framework}\label{sec:inputoutput}

We use input-output theory \cite{CollGard84,GardColl85} to describe the interaction between a quantum system and an external field in terms of the incoming and outgoing field operators. As we need to model a network of connected quantum systems, we use a generalization~\cite{GougJame09,GougJame09a} of the cascaded quantum systems formalism \cite{Carm93,Gard93}. The generalization is called the ``SLH'' framework ~\cite{CombKercSaro17}.

The SLH formalism assigns to each site of a network an operator triple $G = (S,L,H)$. The $L$ operator couples the field to the local system (e.g.\ $\sqrt{\gamma} a$ is the operator that couples a single mode cavity to the continuum), while $H$ is the local system Hamiltonian (e.g.\ $\Delta a\dg a$). The $S$ operators is trivial for all systems in this paper, so from now on we drop it {(and refer simply to ``LH'' parameters)}. The SLH formalism then provides a set of algebraic rules to obtain the parameters for the entire network in terms of the individual sites, and subsequently the collective Heisenberg equations of motion and input-output relations. 

Next we leverage the relationship between input-output theory and the scattering formalism \cite{DaltBarnKnig99,FanKocaShen10} to compute the S-matrices that describe one and two photon transport in the systems of \cref{fig:nsites}.  For a detailed account on the relationship between the input-output and scattering formalisms, see Refs.~\cite{GardColl85,DaltBarnKnig99,FanKocaShen10,PletGrit12,RoyWilsFirs16,BrodComb16a}.

The external field is described by input and output operators $\ain(t)$ and $\aout(t)$ satisfying
\begin{equation}
\left[\ain(t) ,\ain\dg(s) \right]=\delta(t-s)=\left[\aout(t) ,\aout\dg(s) \right].
\end{equation}
We can define analogously delta-commuting frequency-domain operators $\ain(\omega)$, related to $\ain(t)$ by
\begin{equation} \label{eq:binFourier}
\ain(t) = \frac{1}{\sqrt{2\pi}}\int d\omega \, \ain (\omega)e^{-i\omega t},
\end{equation}
with an equivalent relation between $\aout(\omega)$ and $\aout(t)$.
We can then define the scattering eigenstates (i.e.\ states which are frequency eigenstates at the asymptotic past/future) as
\begin{align}
\ain\dg(\omega)\ket{0} = \ket{\omega^+}\quad {\rm and} \quad \aout\dg(\omega)\ket{0} = \ket{\omega^-},
\end{align}
and the local system is assumed to be in the ground state.
From the scattering states we define the single-photon S-matrix:
\begin{align}
S_{\omega  ,\nu } & \defeq \expt{\omega^-|\nu^+}= \expt{0|\aout(\omega)\ain\dg(\nu)|0}. \label{eq:1scat}
\end{align}
The two-photon S-matrix can be written analogously as
\begin{align}
S_{\omega_1,\omega_2,\nu_1,\nu_2} & \defeq \braket{\omega_1 ^{-} \omega_2^{-}|\nu_1^+ \nu_2^+}   \nonumber\\
& =  \bra{0} \aout(\omega_1 )\aout(\omega_2 )\opin{a\dg}(\nu_1)\opin{a\dg}(\nu_2)\ket{0}. \label{eq:2scat}
\end{align}

Propagating photons are not single-frequency entities, they must be described by wavepackets. The final ingredient needed is to specify an input photon in some frequency-domain wave packet $\ket{1_\xi}=\int d\nu \, \xi(\nu) \opin{a^{\dagger}}(\nu) \ket{0}$. Then we can use the S-matrices to obtain the output wavepacket as, e.g.,
\begin{align*}
\ket{1_{\tilde{\xi}}} & = \int d\nu \, d\omega \, S_{\omega ,\nu } \xi(\nu) \opout{a^{\dagger}}(\omega) \ket{0} \\
& = \int d\omega \, \tilde{\xi}(\omega) \opout{a^{\dagger}}(\omega) \ket{0},
\end{align*}
where $\tilde{\xi}(\omega) = \int d\nu \,  S_{\omega ,\nu } \xi(\nu)  $, with an analogous description for two-photon transport.

\section{Single-site Self-Kerr scattering}\label{sec:single}

Refs.~\cite{LiaoLaw10,XuRephFan13} derived the S-matrix for one- and two-photon transport through a single-site self-Kerr medium inside a cavity. Using our notation and 
formalism, the system considered in \cite{LiaoLaw10,XuRephFan13} can be described by the LH parameters:
\begin{align}
G_{\rm sys} 
&=\left(\sqrt{\gamma} a, \Delta a\dg a + {\chi a\dg a a\dg a} \right), \label{eq:XuSLH}
\end{align}
with the cavity decay rate $\gamma$, resonance frequency $\Delta$, and interaction strength $\chi$. 
The S-matrix obtained in \cite{LiaoLaw10,XuRephFan13} can be written as in \crefrange{eq:singleSdelta}{eq:twoSdeltaT} with
\begin{equation}
t_{\omega} = \barG{\omega}, \label{eq:1photonSmatknown}
\end{equation}
and 
\begin{align} \label{eq:2photonSmatknown}
C_{\omega_1,\omega_2,\nu_1,\nu_2} & = - \frac{\chi \gamma^2}{2\pi} \left( 1+\frac{i \chi}{\Gam{\omega_1}+\Gam{\omega_2}} \right)^{-1} \nonumber\\
& \hphantom{-} \times  \frac{1}{\Gam{\nu_2}\Gam{\nu_1} \Gam{\omega_2} \Gam{\omega_1}},
\end{align}
where we defined the shorthands
\begin{align} \label{eq:shorthands}
\Gam{\omega}\defeq\frac{\gamma}{2}+i(\Delta-\omega),\quad 
\barG{\omega}\defeq \frac{\conG{\omega}}{\Gam{\omega}}.
\end{align}
This result can be written in the original notation of \cite{LiaoLaw10,XuRephFan13} by setting $\chi \rightarrow \chi/2$, $\gamma \rightarrow 2/\tau$, and $\Delta \rightarrow \omega_c$.

Next we show how a similar S matrix can be obtained via a cross-Kerr interaction supplemented by a mirror, most notably in terms of the spectral entanglement of outgoing wave packets. However, as we concatenate these interaction sites into larger networks, different behaviors will emerge.

\subsection{From a cavity-mediated cross-Kerr interaction to a self-Kerr interaction}\label{sec:crossker}
A cavity-mediated cross-Kerr interaction site is defined by the LH parameters
\begin{align}
G
=\left(\left[\begin{array}{c}\sqrt{\gamma} a \\\sqrt{\gamma} b\end{array}\right ], \Delta a\dg a +\Delta b\dg b +\chi a\dg a b\dg b \right). \label{eq:cavitySLH}
\end{align}
As there are two entries in the L parameter, this is a two-input and two-output system.

\begin{figure}[ht]
\includegraphics[width=\columnwidth]{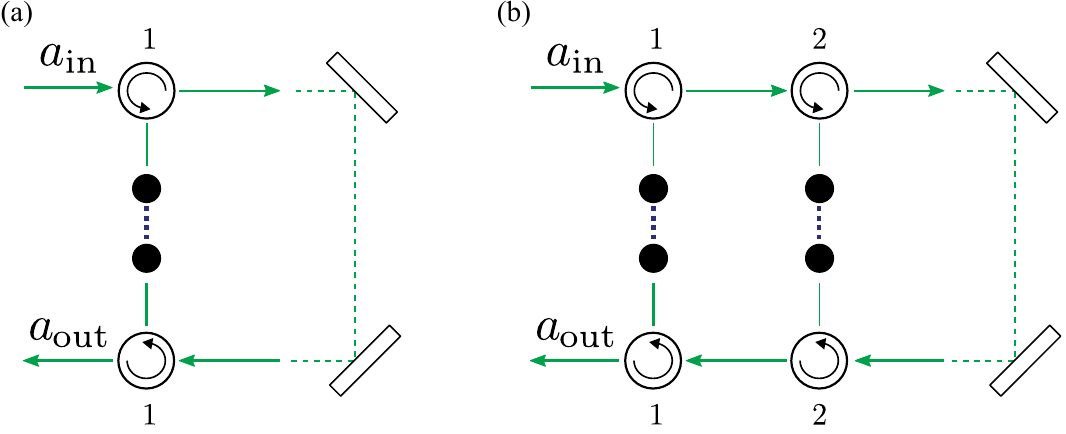}
\caption{(Color Online). (a) A single cross-Kerr interaction site which we supplement with a mirrors to mimic a self-Kerr site. (b) The generalization of (a) to a chain with two sites.}\label{fig:12sites}
\end{figure}

We now consider a mirror that routes the photons from the output of port $a$ to the input of port $b$, as in \cref{fig:12sites}(a). This is accomplished mathematically by using the feedback reduction rule (see rule 4 in \cite{CombKercSaro17}), which lead to the new LH parameters
\begin{subequations} \label{eq:SLH1}
\begin{align} 
L_1 =&  \sqrt{\gamma} (a + b), \label{eq:L1} \\
H_1 =& \Delta (a\dg a + b\dg b) + \chi a\dg a b\dg b + \frac{\gamma}{2i} (b\dg  a - a\dg b). \label{eq:H1}
\end{align}
\end{subequations}
The subscripts on L and H indicate these parameters are for a single site with one input and one output.

The LH parameters  give rise to the input-output relation
\begin{align} 
\aout (t) &= \sqrt{\gamma} \left(a + b\right) + \ain (t), \label{eq:io1site}
\end{align}
and corresponding equations of motion for the cavity operators \cite{CombKercSaro17}
\begin{subequations}\label{eq:DE1}
\begin{align}
\partial_t a &= -\left(\frac{\gamma}{2}+i \Delta\right)a - i \chi a b \dg b - \sqrt{\gamma} \ain(t) \label{eq:DE1a} \\
\partial_t b &= -\left(\frac{\gamma}{2}+i \Delta\right)b - i \chi a\dg a b - \gamma a - \sqrt{\gamma} \ain(t). \label{eq:DE1b}
\end{align}
\end{subequations}

\begin{figure}[h]
\includegraphics[width=0.8\columnwidth]{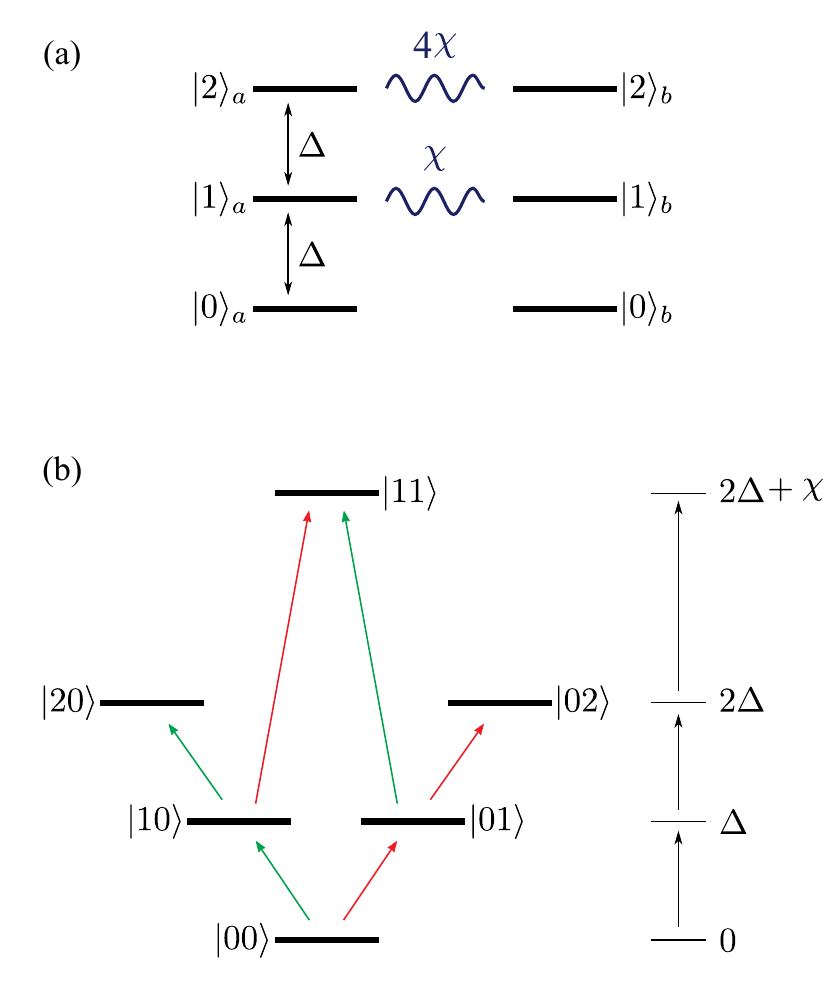}
\caption{(Color Online). (a) A realization of our interaction site using two three-level atoms. The transitions of each atom ($a$ and $b$) are addressed by different field modes. The level structure is constructed so as to only induce a nonlinear effect if both photons are absorbed, one by each atom. The component proportional to $4\chi$ is never accessed by two photons, and so can be dropped from the level structure. (b) A 6-level structure equivalent to (a) if there are less than two excitations in the field. Each transition is addressed by a different field mode, colored as in \cref{fig:cavities}(a).}\label{fig:atoms}
\end{figure}

These equations of motion were obtained assuming that the interaction is mediated by a cross-Kerr medium inside a cavity, but analogous equations could be obtained using atomic systems. In \cref{fig:atoms} we show two atomic realizations of our proposal. Through the rest of the paper we restrict our description to the cavity case, but the final results would be the same in all these systems as long as we consider at most two-photon scattering.

\subsection{Single-photon S-matrix}\label{sec:1site1photon}
Using \cref{eq:io1site} and the definition of the one photon S-matrix in \cref{eq:1scat}, we can write
\begin{align} \label{eq:SM1p1s}
S_{\omega, \nu} 
&= \frac{1}{\sqrt{2\pi}} \int dt \bra{0} \aout (t) \ket{\nu^{+}} e^{i \omega  t} \notag \\
&= \delta(\omega  - \nu) + \sqrt{\frac{\gamma}{2\pi}} \int dt \bra{0} a + b \ket{\nu^{+}} e^{i \omega  t}.
\end{align}

Taking the relevant matrix elements of \cref{eq:DE1a} and solving the resulting differential equation, we obtain the single-photon S-matrix
\begin{align} 
S_{\omega, \nu} =  \barG{\omega}^2 \delta(\omega-\nu) \label{eq:SM1p1sres}
\end{align}
where we used the identity
\begin{align}
\barG{\omega} = \frac{\gamma}{\Gam{\omega}}-1.
\end{align}
Note that the S-matrix in \cref{eq:SM1p1sres} has the form of \cref{eq:singleSdelta}, with $t_{\omega} = \barG{\omega}^2$. Comparing with \cref{eq:1photonSmatknown}, we see that \cref{eq:SM1p1sres} has twice the phase $\barG{\omega}$. This is because in our system of \cref{fig:cavities}(b) the photons are effectively scattered by two cavities.

\subsection{Two-photon S-matrix}\label{sec:1site2photon}
Let us now find the two-photon S matrix. In this Section we use the techniques developed in Appendix A of Ref.~\cite{BrodComb16a}. We begin by introducing a resolution of the identity between $\aout(\omega_1 )$ and $\aout(\omega_2)$ in \cref{eq:2scat}. Due to photon number conservation, the identity operator on the relevant subspace is $\int ds \ket{s^{+}}\bra{s^{+}}$. Using \cref{eq:io1site} and \cref{eq:SM1p1sres} we write
\begin{align*}
S_{\omega_1,\omega_2,\nu_1,\nu_2} =&  \barG{\omega_1}^2 \bigg[  \delta(\omega_1 - \nu_1) \delta(\omega_2 - \nu_2)\nonumber\\
& \hphantom{\barG{\omega_1}^2 \bigg[} + \delta(\omega_1 - \nu_2) \delta(\omega_2 - \nu_1) \\ 
& \hphantom{\barG{\omega_1}^2 \bigg[} + \sqrt{\frac{\gamma}{2\pi}} \int dt e^{i \omega_2 t} \bra{\omega_1 ^+} a+b \ket{\nu_1^+ \nu_2^+}\! \bigg]
\end{align*}
This is similar to Eq.\ (A6) of \cite{BrodComb16a}, except that now we have the sum of two delta functions, which is simply the manifestation of bosonic statistics. 
Solving the remaining part of the S matrix  results in 
\begin{align}
C_{\omega_1,\omega_2,\nu_1,\nu_2} = & - \frac{\chi \gamma^2}{2\pi} \left( 1+\frac{i \chi}{\Gam{\omega_1}+\Gam{\omega_2}} \right)^{-1} \nonumber\\
& \hphantom{-} \times  \frac{\left(\barG{\nu_1}+\barG{\nu_2} \right)\left(\barG{\omega_1}+\barG{\omega_2} \right)}{\Gam{\nu_2}\Gam{\nu_1} \Gam{\omega_2} \Gam{\omega_1}},\label{eq:2p1sres}
\end{align}
which has the same form as \cref{eq:2photonSmatknown}.
$C_{\omega_1,\omega_2,\nu_1,\nu_2}$ encodes the interaction term of the S matrix. 
The $\left(\barG{\nu_1}+\barG{\nu_2} \right)\left(\barG{\omega_1}+\barG{\omega_2} \right)$ factor is interpreted as follows. In the scattering channel where the photons interact, each must be absorbed by a different cavity mode, since this is the only source of nonlinearity by design.
The photon absorbed by mode $a$ during the interaction also picks up a linear phase at mode $b$, leading to the $\barG{\nu_1} + \barG{\nu_2}$ factor (note the $\nu_1 \leftrightarrow \nu_2$ symmetry). Similarly, the $\barG{\omega_1} + \barG{\omega_2}$ term comes from the linear phase that the other photon picks up at mode $a$. From here on, for simplicity of notation we indicate the $\nu_1 \leftrightarrow \nu_2$ symmetry explicitly in equations.

Combined, \cref{sec:crossker}, \cref{sec:1site1photon}, and this Section show that, at the two-photon level, a self-Kerr interaction can be simulated by a cross-Kerr interaction and a mirror.

\section{Two-site  self-Kerr scattering}\label{sec:2counter}

We now extend the results of \cref{sec:single} to two interaction sites. The physical system we consider is represented in \cref{fig:12sites}(b). 
The LH parameters, after doing the feedback connection, are 
\begin{subequations}
\begin{align}
L &= \sqrt{\gamma} (a_1 + a_2 +b_1 + b_2)\\
H_T &= H_{\rm self} + H_{\rm int} + H_{\rm cas}
\end{align}
\end{subequations}
where the Hamiltonian has an on-site contribution coming from the atoms' self energies and an interaction term, plus an effective Hamiltonian arising from cascading:
\begin{subequations}
\begin{align}
H_{\rm self} &= \Delta (a_1 \dg a_1+a_2\dg a_2 + b_1\dg b_1+ b_2\dg b_2 )\\
H_{\rm int} &= \chi (a_1 \dg a_1 b_1 \dg b_1+a_2 \dg a_2 b_2 \dg b_2)\\
H_{\rm cas} &=\frac{\gamma}{2i} (b_1\dg a_1 + b_1\dg a_2 + b_1\dg b_2 + b_2\dg a_1 +\nonumber \\
&\quad \quad\quad b_2\dg a_2 + a_2 \dg a_1 + {\rm H.c.} ).
\end{align}
\end{subequations}

The corresponding two-site input-output relation and equations of motion are:
\begin{subequations} \label{eq:inputoutput2site}
\begin{align} 
\aout (t) =& \sqrt{\gamma} \left(a_1 + a_2 + b_1 + b_2\right) + \ain (t), \label{eq:io2site}\\
\partial_t a_1 =& -\left(\frac{\gamma}{2}+i \Delta\right)a_1 - i \chi a_1 b_1\dg b_1 - \sqrt{\gamma} \ain(t),\label{eq:DE2sitea}\\
\partial_t a_2 =& -\left(\frac{\gamma}{2}+i \Delta\right)a_2 - i \chi a_1 b_2\dg b_2 - \sqrt{\gamma} \ain(t),\nonumber\\
& - \gamma a_1\label{eq:DE2siteb} \\
\partial_t b_2 =& -\left(\frac{\gamma}{2}+i \Delta\right)b_2 - i \chi a_2\dg a_2 b_2 - \sqrt{\gamma} \ain(t),\nonumber\\
& - \gamma (a_1+a_2) \label{eq:DE2sitec}\\
\partial_t b_1 =& -\left(\frac{\gamma}{2}+i \Delta\right)b_1 - i \chi a_1\dg a_1 b_1 - \sqrt{\gamma} \ain(t)\nonumber\\
& - \gamma (a_1+a_2+b_1) .\label{eq:DE2sited}
\end{align}
\end{subequations}
Since from here on many calculations involved in obtaining the S-matrices are similar the procedure from \crefrange{sec:1site1photon}{sec:1site2photon} and Ref.~\cite{BrodComb16a}, we simply skip the details and focus on the interpretation of the S-matrices.\\

\subsection{Single-photon S-matrix}

The single-photon S-matrix for two sites is:
\begin{align}  \label{eq:SM1p2s}
S_{\omega, \nu} 
=& \delta(\omega  - \nu) +\nonumber\\
 &\sqrt{\frac{\gamma}{2\pi}} \int dt \bra{0} a_1 + a_2 + b_1 + b_2 \ket{\nu^{+}} e^{i \omega  t}.
\end{align}
From this, the resulting single-photon S-matrix can be written as in \cref{eq:singleSdelta}, where
\begin{align} 
t_{\omega} = \barG{\omega}^4 \label{eq:SM1p2sres}
\end{align}
is the single-photon phase, implying four scattering events. 

\subsection{Two-photon S-matrix}
Following the previous steps we write the two-photon S-matrix  as
\begin{widetext}
\begin{align*}
S_{\omega_1,\omega_2,\nu_1,\nu_2} =  \barG{\omega_1}^4 \bigg[  \delta(\omega_1 - \nu_1) \delta(\omega_2 - \nu_2) +
\delta(\omega_1 - \nu_2) \delta(\omega_2 - \nu_1)+ \sqrt{\frac{\gamma}{2\pi}} \int dt e^{i \omega_2 t} \bra{\omega_1 ^+} a_1 + a_2 + b_1 + b_2 \ket{\nu_1^+ \nu_2^+} \bigg].
\end{align*}
Which, together with \crefrange{eq:io2site}{eq:DE2sited}, leads to a two-site, two-photon S-matrix with the form of \cref{eq:twoSdelta} where

\begin{align}
C_{\omega_1,\omega_2,\nu_1,\nu_2} = & - \frac{\chi \gamma^2}{2\pi} \left( 1+\frac{i \chi}{\Gam{\omega_1}+\Gam{\omega_2}} \right)^{-1} \frac{  
\left(\barG{\nu_1}^3+\barG{\nu_2}^3 \right)\left(\barG{\omega_1}^3+\barG{\omega_2}^3\right)
+\barG{\nu_1}\barG{\nu_2}\barG{\omega_1}\barG{\omega_2}\left(\barG{\nu_1}+\barG{\nu_2} \right)\left(\barG{\omega_1}+\barG{\omega_2}\right)
}
{\Gam{\nu_2}\Gam{\nu_1} \Gam{\omega_2} \Gam{\omega_1}}
\label{eq:2p2sres}
\end{align}
\end{widetext}

This expression can be interpreted similarly as \cref{eq:2p1sres}. The nonlinearity is the sum of two contributions. The first, proportional to $\left(\barG{\nu_1}^3+\barG{\nu_2}^3 \right)\left(\barG{\omega_1}^3+\barG{\omega_2}^3\right)$, corresponds to the case where photons interact at site $1$. The sum of phases encodes phases picked up by photons in those sites where they did not interact. One photon picks up phase $\barG{\omega_1}^3$ or $\barG{\omega_2}^3$ while interacting with the cavity modes $a_1$, $a_2$ and $b_2$, while the other phase is picked up by the other photon at modes $a_2$, $b_2$ and $b_1$. Similarly, the second term corresponds to the case where photons interacted at site 2. Both photons interact with mode $a_1$ and $b_1$, hence the phases $\barG{\omega_1}\barG{\omega_2}$ and $\barG{\nu_1}\barG{\nu_2}$ respectively. Finally, one input photon picked up a phase from mode $a_2$, and one output photon picked up a phase from mode $b_2$.

This S-matrix has the interesting property of not containing terms with nonlinearities coming from more than one site. This is analogous to what was observed for counter-propagating photons interacting via cross-Kerr interaction sites~\cite{MassFlei04}, as in \cite{BrodComb16a,BrodComb16b}. There, the fact that photons were propagating along the chain in opposite directions meant that, due to causality, they could only interact at a single site. As the number of sites increased, this was interpreted as responsible for the vanishing of spectral entanglement between output photons, and ultimately for the good performance of this system as a two-photon CPHASE gate.

For self-Kerr interactions, counter-propagating conditions are effectively mimicked by the introduction of the mirror in \cref{fig:12sites}(b). Clearly, causality precludes the photons from interacting in more than one site: when they interact at site $i$ it is because one was absorbed into mode $a_i$ and the other into mode $b_i$. But afterwards the photons continue to propagate in different directions, and do not meet again at another site. This suggests we have successfully emulated the counterpropagating conditions that led to the high fidelity CHPASE gate. In the unwrapped lattice {[see \cref{fig:nsites}(b)]}, the lack of interactions at {more than one site} is also enforced by causality, {since it translates to the impossibility of photons propagating to the opposite side of the lattice instantaneously}. Let us now confirm this by analyzing this system's behavior for larger numbers of sites.

\section{\texorpdfstring{$N$}{N}-site scattering and the continuum limit}\label{sec:Ncounter}
We now consider the scattering problem on a chain of $N$ interaction sites, as in \cref{fig:nsites}. In Refs.\ \cite{BrodComb16a,BrodComb16b}, we formulated the  $N$-site problem and determined the S-matrix by induction. The current problem requires a similar calculation which would be a cumbersome reapplication of the same steps from \crefrange{sec:single}{sec:2counter} and those in \cite{BrodComb16a,BrodComb16b}. Instead, we simply conjecture a form for the $N$-site S-matrix, based on prior calculations and our interpretations of \cref{eq:2p1sres} and \cref{eq:2p2sres}.

\subsection{Conjectured one and two-photon S matrices}
We conjecture the single-photon S matrix to be
\begin{align} 
S_{\omega, \nu} = \barG{\omega}^{2N} \delta(\omega-\nu). \label{eq:SM1pNsres}
\end{align}
This is the immediate generalization of \cref{eq:SM1p1sres} and \cref{eq:SM1p2sres} where a photon picks up phases from each of the $2N$ cavity modes it couples to.

Our conjecture for the two-photon S matrix is of the form of \cref{eq:twoSdelta}, with
\begin{equation}
t_\omega = \barG{\omega}^{2N},
\end{equation}
as per \cref{eq:SM1pNsres},  and 
\begin{align}
C_{\omega_1,\omega_2,\nu_1,\nu_2} = & - \frac{\chi \gamma^2}{2\pi} \left( 1+\frac{i \chi}{\Gam{\omega_1}+\Gam{\omega_2}} \right)^{-1} \frac{1}{\Gam{\nu_2}\Gam{\nu_1} \Gam{\omega_2} \Gam{\omega_1}} \nonumber\\
& \hphantom{-} \times \sum_{j=1}^{N}\left(\barG{\omega_1}\barG{\omega_2}\barG{\nu_1}\barG{\nu_1}\right)^{N-j} \nonumber \\
& \hphantom{- \times \sum_{j=1}^{N}} \left(\barG{\omega_1}^{2j-1}+\barG{\omega_2}^{2j-1} \right)\left(\barG{\nu_1}^{2j-1}+\barG{\nu_2}^{2j-1} \right).
\label{eq:2pNsfinal}
\end{align}
This is the natural generalization of previous results to the $N$-site case. It has the same nonlinear term as in \cref{eq:2p2sres}, which is multiplied by a summation over phases. Each term $j$ in this summation comes from the scattering channel where photons interact at site $N+1-j$, and the summand correspond to the linear phases picked up in those sites where the photons did not interact. As an additional check that this expression is sensible, one can easily re-obtain \cref{eq:2p1sres} and \cref{eq:2p2sres} for the $N=1$ and $N=2$ cases respectively. By using the identity $\sum_{i=1}^{N} x^{n-i}y^{i-1} = ( x^n-y^n ) / ( x-y )$ 
we can write \cref{eq:2pNsfinal} in a simpler form
\begin{align}
C_{\omega_1,\omega_2,\nu_1,\nu_2} = & - \frac{\chi \gamma^2}{2\pi} \left( 1+\frac{i \chi}{\Gam{\omega_1}+\Gam{\omega_2}} \right)^{-1} \frac{1}{\Gam{\nu_2}\Gam{\nu_1} \Gam{\omega_2} \Gam{\omega_1}}\notag\\
& \times \left[ \frac{(\barG{\omega_2}\barG{\nu_2})^{2N}-(\barG{\omega_1}\barG{\nu_1})^{2N}}{(\barG{\omega_2}\barG{\nu_2})-(\barG{\omega_1}\barG{\nu_1})}\right] + \nu_1 \leftrightarrow \nu_2.
\label{eq:2pNsfinalsummed}
\end{align}

\subsection{Continuum limit} \label{sec:infsite}

From the conjectured form of the S-matrix in \cref{eq:2pNsfinalsummed} we can also look at the behavior of our system at the $N \rightarrow \infty$ limit. We begin by rewriting the interference term
\begin{align}
  \mathcal{I} :=& \frac{(\barG{\omega_2}\barG{\nu_2})^{2N}-(\barG{\omega_1}\barG{\nu_1})^{2N}}{(\barG{\omega_2}\barG{\nu_2})-(\barG{\omega_1}\barG{\nu_1})} \\
 = &e^{(2N-1)(\phi_1 + \phi_2)}\frac{\sin \left[2N(\phi_1-\phi_2)\right]}{\sin(\phi_1-\phi_2)} \label{eq:prelim}
\end{align}
where we defined $\phi_i$ such that $\barG{\omega_i}\barG{\nu_i} = e^{2 i \phi_i}$ for $i=1,2$ (note from definition that $\barG{\omega}$  is simply a phase). Now we make two crucial assumptions: first, that wave packets are spectrally narrow, and second that they are on-resonance with the cavities, such that all relevant frequencies are sufficiently concentrated around $\Delta$. This allows us to write
\begin{equation}
\phi_i \approx \frac{2}{\gamma} (\omega_i + \nu_i - 2 \Delta)
\end{equation}
for $i=\{1,2\}$. This, together with the fact that $\lim_{N \rightarrow \infty} \sin{N x}/x = \pi \delta(x)$, leads to
\begin{equation*}
\mathcal{I} = \frac{\gamma \pi}{2} e^{2(2N-1)\phi_1} \delta(\omega_1 +\nu_1-\omega_2-\nu_2).
\end{equation*}
The corrections to the width of this delta function are of order $1/N$, which implies the incident photon must have a spectral bandwidth of order $\gamma/N$.
The same reasoning holds for the term where the roles of $\nu_1$ and $\nu_2$ are reversed. By using the approximation above and
\begin{equation*}
\delta(\omega_1+\omega_2-\nu_1-\nu_2)\delta(\omega_1+\nu_1-\omega_2-\nu_2)=\frac{1}{2}\delta(\omega_1-\nu_2)\delta(\omega_2-\nu_1)
\end{equation*}
we rewrite the S matrix as
\begin{align}
&S_{\omega_1,\omega_2,\nu_1,\nu_2}\nonumber\\
&=\barG{\omega_1}^{2N}\barG{\omega_2}^{2N} 
\left[\delta(\omega_1 - \nu_1) \delta(\omega_2 - \nu_2) + \delta(\omega_1 - \nu_2) \delta(\omega_2 - \nu_1)\right] \notag \\ 
&\quad\times 
\left[1 -i\frac{\chi \gamma^3}{8} \left( 1+\frac{i \chi}{\Gam{\omega_1}+\Gam{\omega_2}} \right)^{-1} \frac{1}{|\Gam{\omega_1}\Gam{\omega_2}|^2}  \right].
\label{eq:2pInfsres}
\end{align}
From this we see that the S-matrix (and T-matrix) has reduced to the form of (\ref{eq:idealSmatrix}). Thus, our previous intuition is manifest: for very long chains the spectral entanglement vanishes, and we approximately circumvent the restriction imposed by the CDP. This is the self-Kerr analogue of the result of \cite{BrodComb16a,BrodComb16b}, and is equal (up to a redefinition of $\chi$) to Eq.\ (66) of \cite{BrodComb16a}. The only difference is in symmetrization with respect to exchange of $\nu_1$ and $\nu_2$. Physically, this term originates from bosonic statistic, since here the photons are identical, whereas in \cite{BrodComb16a} they could be distinguished by the chiral modes in which they propagated. In any case, the effective counter-propagation leads to momentum and energy conservation becoming independent conditions, thus removing spectral entanglement.

An S-matrix equivalent to the ideal two-photon self-Kerr effect can be obtained if we further simplify \cref{eq:2pInfsres} by making the approximation $\Gam{\omega} \approx \gamma/2$, in which case the term in the square brackets becomes the phase 
\begin{align}
\left (\frac{\gamma -i \chi}{\gamma +i \chi} \right )&={e^{-i2\tan^{-1}(\chi/\gamma)}}= e^{-i \Phi(\chi,\gamma)},
\end{align}
which should be compared with $C_{\nu_1,\nu_2}$ from  \cref{eq:idealSmatrix}. In the $\chi \rightarrow \infty$ limit $ \Phi(\chi,\gamma) = \pi$  so that
\begin{align}
&\braket{\omega_1^- \omega_2^-|\nu_1^+ \nu_2^+} \nonumber \\
&=-\barG{\omega_1}^{2N}\barG{\omega_2}^{2N} \left[\delta(\omega_1 - \nu_1) \delta(\omega_2 - \nu_2) + \delta(\omega_1 - \nu_2) \delta(\omega_2 - \nu_1)\right].
\label{eq:2pInfsIdeal}
\end{align}
This S-matrix indicates that, up to bosonic statistics and linear phase effects, the two-photon wavepacket simply acquires a $\pi$ phase, as desired.

\section{Interpretation and discussion of the CDP}\label{sec:CDP}
\begin{figure}[ht]
\includegraphics[width=0.9\columnwidth]{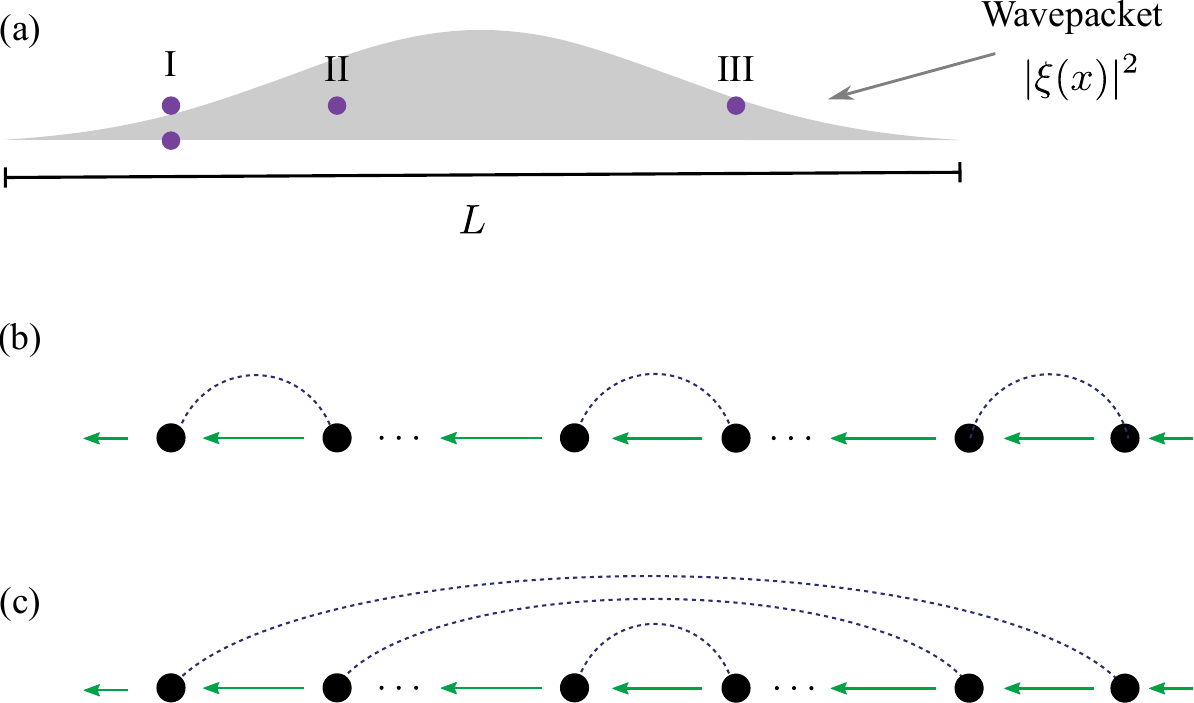}
\caption{(Color Online). (a) We interpret the square of the wavepacket as the probability to find the photon in a given position $x$. Consider the semiclassical picture where we sample photons from the wavepacket at discrete sites. Suppose the first photon is sampled at some position, and the second at different distances from it (I - III). If interactions are local, as in (b), only when the second photon is at position (II) do they interact. With interactions of varying distances, as in (c), the photons are more likely to interact no matter how far apart they were sampled. The non-locality structure of (c) is effectively mimicked by our mirror arrangement [cf.\ \cref{fig:nsites}].
 }\label{fig:interp}
\end{figure}

Let us discuss one intuitive way to understand our results and how they relate to the cluster decomposition principle. To that end, consider a semiclassical picture where we suppose the initial position of the two photons are sampled from their (common) wavepacket distribution $|\xi(x)|^2$ far away from the interaction region, see \cref{fig:interp}. We assume the wavepacket has spatial (or temporal) length $L$. The two particles then propagate towards the interaction region with the sampled distance between them.

In our proposal, the interaction region consists of discrete scattering sites that interact according to some connectivity graph. The locality of the graph essentially determines the probability the particles will interact. From \cref{fig:interp} we see that, for the particles to be likely to interact independently of how far apart they were sampled, the interaction graph requires non-local connections of all lengths up to order $L$. This range of non-local connections can be effectively achieved via the use of the mirror [cf.\ \cref{fig:nsites}], via counter-propagation \cite{BrodComb16b,BrodComb16a}, or propagation with different group velocities.

One might ask why not use short wavepackets, thus ensuring the photons are likely to be sampled close to each other in the first place. The reason can be traced to the interaction between the field and the atoms: if photons are too broad in the frequency domain, they are not likely to be absorbed by the medium, and so never get the chance to interact. This effect is not captured by our semiclassical picture, but is straightforward to see in the full quantum description.

We have partially answered why many sites are required in the semiclassical picture: so particles sampled from long wave packets have the chance to interact. If we adopt a full wave-like picture, additional interpretation can be given. Specifically, the long range interactions and number of sites ensure the wave packet gets a uniform phase shift. Colloquially, every part of the wavepacket sees all other parts.

Finally, consider the cluster decomposition principle. For any finite size of the medium, if we consider the two photons sampled farther apart than that there would be no interaction, agreeing with one formulation of the CDP \cite{XuRephFan13}. However, as the length of the wavepackets (and the effective medium length) tend to infinity the medium becomes infinitely long, and the interactions effectively extremely nonlocal [if represented in the unwound manner of \cref{fig:nsites}(b)], which formally breaks the assumptions of the CDP.

\section{Application to two-qubit gates}\label{sec:gates}

\begin{figure}[ht]
\includegraphics[width=0.75\columnwidth]{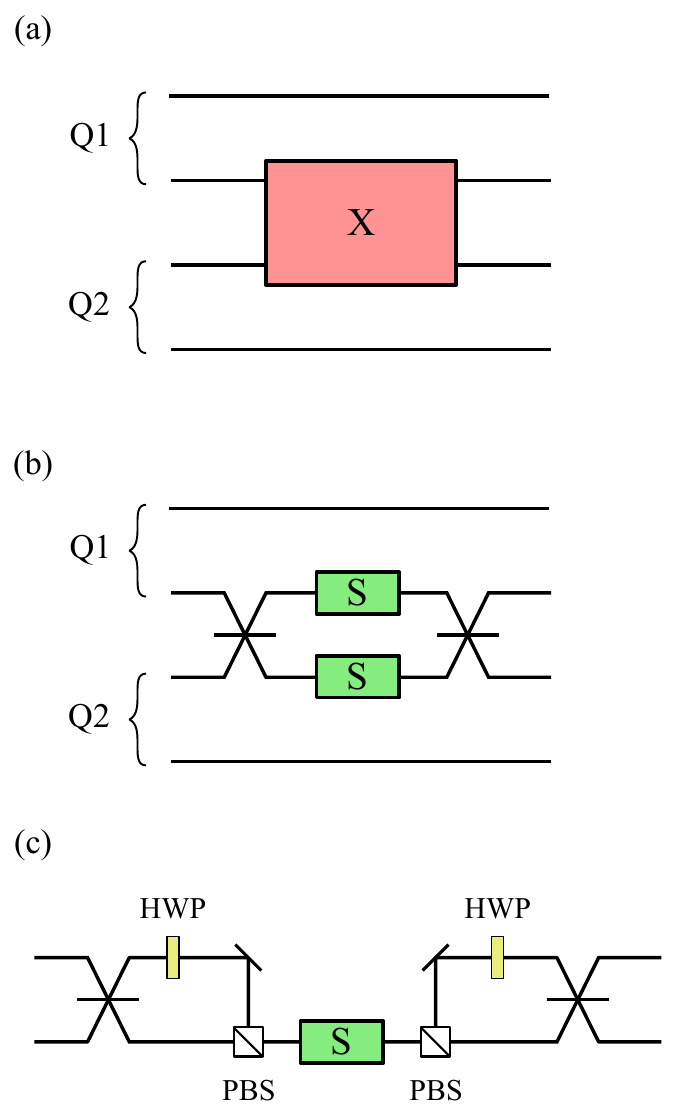}
\caption{(Color Online). Three CPHASE gates using Kerr nonlinearities and dual-rail encoding \cite{KnilLaflMilb01}. For simplicity, we chose the encoding such that the $\ket{1}_{Q1}$ and $\ket{1}_{Q2}$ logical states are given by the $\ket{01}_{12}$ and $\ket{10}_{34}$ physical states respectively. (a) A cross-Kerr interaction applied to one mode of each qubit (first proposed in \cite{Milb89,ChuaYama95}).
(b) Two modes pass through a balanced beam splitter. Whenever both are occupied with a single photon the photons exit together, and are routed to two parallel nonlinear sign-shift gates \cite{KnilLaflMilb01,NystMcCuHeuc17}, which can be implemented using our self-Kerr construction. (c) By using half-wave plates (HWP) and polarizing beam splitters (PBS), we can store the dual-rail encoding temporarily in the photon polarization, enabling the gate with a single self-Kerr interaction. Of course, under the caveat that the interaction must be fully polarization-independent.}\label{fig:CPHASE}
\end{figure}

In this Section we discuss the application of our results to quantum computing.  Many of the original proposals for photon-photon gates were based on cross-Kerr nonlinearities~ \cite{Milb89,ChuaYama95}, see \cref{fig:CPHASE}(a). However, by using a beamsplitter and taking advantage of the Hong-Ou-Mandel effect we can construct a photon-photon gate out of two self-Kerr nonlinearities~\cite{KnilLaflMilb01,NystMcCuHeuc17}, as in \cref{fig:CPHASE}(b).

We have shown that the S-matrix in \cref{eq:2pNsfinalsummed} reduces to the ideal two-photon self-Kerr S-matrix in the limit of long chains ($N\rightarrow \infty$), very large $\chi$ and spectrally narrow on-resonance wave packets. But we can also ask how the chain of \cref{fig:nsites} performs for finite $N$, to gauge whether this proposal could be applicable in practice. 

To that end, let us compare \cref{eq:2pNsfinalsummed} 
with the equivalent $C_{\omega_1,\omega_2,\nu_1,\nu_2}$ obtained from Eq.\ (60) of Ref.~\cite{BrodComb16a}, the result for the $N$-site cross-Kerr setup which we write as (redefining $\chi$ by a factor of 2 to simplify comparison)
\begin{align*}
C_{\omega_1,\omega_2,\nu_1,\nu_2} = & - \frac{\chi \gamma^2}{2\pi} \left( 1+\frac{i \chi}{\Gam{\omega_1}+\Gam{\omega_2}} \right)^{-1} \frac{1}{\Gam{\nu_2}\Gam{\nu_1} \Gam{\omega_2} \Gam{\omega_1}}\notag\\
& \hphantom{-} \times \left[ \frac{(\barG{\omega_2}\barG{\nu_2})^{N}-(\barG{\omega_1}\barG{\nu_1})^{N}}{(\barG{\omega_2}\barG{\nu_2})-(\barG{\omega_1}\barG{\nu_1})}\right]
\end{align*}
Up to the symmetrization on the $\nu$'s, the interaction term in the self-Kerr S-matrix of \cref{eq:2pNsfinalsummed} for $N$ sites corresponds exactly to the interaction term in the cross-Kerr S-matrix of \cite{BrodComb16a} for $2N$ sites. 

The fact that the self-Kerr S-matrix matches the cross-Kerr S-matrix with twice the number of sites can be easily understood. In our  proposals, the S matrix shows that photons actually acquire a large phase shift from a single site (in contrast with, e.g.\ \cite{ChudChuaShap13}, where many sites are required to build up the phase shift). The $N \rightarrow \infty$ limit is necessary for two reasons ~\cite{BrodComb16a,BrodComb16b}. First, to increase the amplitude that photons interact in at least one site. Second, to remove spectral entanglement by interference of different frequency components. In this sense, the chain in \cref{fig:nsites} is equivalent to a chain with twice as many sites as that of \cite{BrodComb16a,BrodComb16b}, since photons effectively transverse $2N$ sites in a round trip inside the chain (even though there are only $N$ couplings between right and left propagating modes).

This does not mean that the setup of \cref{fig:nsites} is twice as efficient as the analogous cross-Kerr one for specific tasks. To see why, consider the standard way to construct a CPHASE gate using these interactions, as in \cref{fig:CPHASE}.

Let qubits be encoded in a standard dual-rail encoding, i.e.\ qubit computational states are $\ket{0}_L= \ket{01}$ and $\ket{1}_L=\ket{10}$, corresponding to a photon that can be in one of two modes (spatial, polarization, or frequency).  An arbitrary single-qubit unitary on this encoding can be achieved with beam splitters and phase shifters. The two-qubit gate we target is the controlled-phase or CPHASE gate. A CPHASE can be achieved with the following transformation in the physical basis
\begin{align}
U(\phi) = \left(\begin{array}{cccc}1 & 0 & 0 & 0 \\0 & 1 & 0 & 0 \\0 & 0 & 1 & 0 \\0 & 0 & 0 & e^{i \phi }\end{array}\right)
\end{align}
when $\phi=\pi$ we have the usual controlled-$Z$ gate. 

It is easy to see that an ideal cross-Kerr interaction between one mode from each qubit, as in \cref{fig:CPHASE}(a), directly implements a CPHASE gate. The standard way \cite{KnilLaflMilb01,NystMcCuHeuc17} to implement the same gate using a self-Kerr interaction is shown in \cref{fig:CPHASE}(b). The Hong-Ou-Mandel effect transforms a $\ket{11}$ state into a superposition of $\ket{02}$ and $\ket{20}$. We route this state to {\bf two} parallel self-Kerr interactions to apply a $\pi$ phase, and finally use a second beamsplitter to recover a $- \ket{11}$ state. We can now replace the each ideal interaction in \cref{fig:CPHASE} by an $N$-site chain of either cross-Kerr \cite{BrodComb16a,BrodComb16b} or self-Kerr (\cref{fig:nsites}) interactions. Since the S-matrices are essentially identical, the self-Kerr chain attains the same gate quality (measured in terms of the average gate fidelity) as a cross-Kerr chain with twice the number of sites.

The similarity between the self-Kerr and cross-Kerr S-matrices already implies a connection between fidelities of any task performed with the present proposal and that of \cite{BrodComb16a,BrodComb16b}, as outlined above. For completeness, in \cref{app:fidelities} we describe how these fidelities can be computed from the reported S-matrices. From these numerical calculations we conclude that, assuming the input photons to be Gaussian wavepackets and our goal to be a CPHASE gate with a $\pi$ phase shift, the best average gate fidelities obtained with our self-Kerr proposal are $95\%, 99\%$ and $99.9\%$ using chains of  $2\times 3$, $2\times 6$ and $2\times 25$ sites, respectively. As the number of sites increases, the input wavepacket that maximizes the fidelity becomes narrower. We can fit the best achievable fidelity $F(\sigma_{\rm opt},N)$ and the corresponding optimal wave packet width $\sigma_{\rm opt}(N)$ to see how they scale as $N$ increases, obtaining~\cite{BrodComb16b}:
\begin{subequations} \label{eq:asympt}
\begin{align}
1-F(\sigma_{\rm opt},N) &=0.537 N^{-1.61},\\
\sigma_{\rm opt}(N)&=0.350 N^{-0.81}.
\end{align}
\end{subequations}

Interestingly, if the interaction sites are polarization-independent, we can reduce the number of sites required by the self-Kerr chain by half. This is depicted in \cref{fig:CPHASE}(c). By using a half-wave plate and a polarizing beam splitter we can map the $\ket{02} + \ket{20}$ state e.g.\ into a $\ket{HH} + \ket{VV}$ state. As long as the medium acts identically on both polarization states, a single self-Kerr chain can induce a $\pi$ phase on both polarization sectors at once. This would reduce the required number of sites for $ 95\%, 99\%$ and $99.9\%$ fidelities to 3, 6 and 25, respectively, at the cost of more stringent constraints on the interaction sites.

On the other hand, if the desired task \cite{StobMilbWodk08,GenoInvePari09,RossAlbaPari16} requires implementing the self-Kerr transformation on a propagating mode, our new proposal might be more efficient. Furthermore, the observation that the self-Kerr chain  only contains $N$ actual points of interaction, but that interference effects happens as if there were $2N$ sites, helps to elucidate the role of interference in obtaining a high-fidelity approximation in the $N\rightarrow \infty$ limit. For example, it raises the question of whether there is a different geometrical arrangement of interactions (see e.g.\ \cite{ThomGoklLloy15}) that maximizes interference effects while minimizing the number of actual nonlinearities. This would be beneficial in practice since the latter are more technologically demanding.

\section{Conclusions}\label{sec:conclusion}
In this paper, we showed how to construct a microscopic model for a medium that approximates the behaviour of a self-Kerr medium in the one- and two-photon regime. The main idea is to use a mirror at the end of a chain of cross-Kerr sites to simulate counterpropagation for two photons that propagate in the same mode, allowing us to draw from intuition developed in previous work \cite{BrodComb16b,BrodComb16a}. We also discuss how our proposal satisfies the cluster decomposition principle for any finite size (i.e.\ finite number of interaction sites), but more importantly we also show how one can get arbitrarily close to violating the CDP as the length of chain increases.

One possible application of this proposal is in the construction of a photonic two-qubit CPHASE gate. We conclude that our proposal can achieve high fidelities (e.g.\ $99\%$) with a modest number of interaction sites (12). The resource count of our self-Kerr proposal seems to be exactly the same as the cross-Kerr construction of \cite{BrodComb16b}, although the self-Kerr proposal might be more flexible under particular circumstances.

With respect to quantum computing applications the main open question is whether the insight on the role of interference discussed in \cref{sec:gates} can be leveraged to propose even more efficient constructions. 
In our self-Kerr proposal the interference effects in the S-matrix accumulate twice as fast as in our cross-Kerr proposal, for the same number of sites. It is these interference effects that are responsible for the decay of spectral entanglement as the size of the chain increases. Therefore, one can imagine a more complex situation where photons traverse an $N$-site chain many times (or using a different interaction network), building up the interference effects faster. The intuition developed here suggests such a proposal might lead e.g.\ to a more efficient CPHASE gate (with respect to number of interaction sites). Other natural open questions include, for example, the analysis of the effect of experimental imperfections such as thermal noise and losses \cite{GorsOtteDeml10}, or the calculation of S matrices for larger number of photons.

One fundamental and open question for nonlinear quantum optics is to relate the two-photon S-matrix we derived to an effective field-field interaction inside some dielectric. 
For example, much of the older quantum optics literature~\cite{DrumCart87,LaiHaus89,Wrig90,AbraCohe94,BoivKartHaus94} considers fields that  interact directly via some Hamiltonian of the type
\begin{align}
\chi \int dx\int dy V(x-y) \phi\dg(x)\phi\dg(y)\phi(y)\phi(x) 
\end{align}
where $\phi\dg(x)$ and $\phi(x)$ are field operators, and $V(x-y)$ is a potential, e.g. $V(x-y)= \delta(x-y)$.  In contrast, our work explicitly models the matter that mediates the field-field interactions.
It would be
interesting to see if the $S$ matrices were reported here could be converted into an effective field-field interaction Hamiltonian \cite{Datt97,Dutr05}.

Upon completion of this work we became aware of recent work by \citet{CoheMolm} which considers a similar physical system. Their final goal is in a sense the inverse of ours: their qubits are encoded in the chain sites (each including a qubit encoded in a three-level atom), and the field is used to mediate the two-qubit gates. In our proposal, the qubits are encoded in the photons and the chain of atoms acts as the interaction mediator.

{\em Acknowledgments:} The authors acknowledge helpful discussions with Marcelo P.\ Almeida, Anushya Chandran, Dirk Englund, Mikkel Heuck, Johannes Otterbach, Tom Stace, Miles Stoudenmire, and Andrew White. 
JC was supported by the Australian Research Council through a Discovery Early Career Researcher Award (DE160100356) and via the Centre of Excellence in Engineered Quantum Systems (EQuS), project number CE170100009. 
DB acknowledges financial support from CAPES (Brazilian Federal
Agency for Support and Evaluation of Graduate Education) within the Ministry of Education of
Brazil. \\

\bibliography{../../scattering_ref}

\begin{thebibliography}{66}%
\makeatletter
\providecommand \@ifxundefined [1]{%
 \@ifx{#1\undefined}
}%
\providecommand \@ifnum [1]{%
 \ifnum #1\expandafter \@firstoftwo
 \else \expandafter \@secondoftwo
 \fi
}%
\providecommand \@ifx [1]{%
 \ifx #1\expandafter \@firstoftwo
 \else \expandafter \@secondoftwo
 \fi
}%
\providecommand \natexlab [1]{#1}%
\providecommand \enquote  [1]{``#1''}%
\providecommand \bibnamefont  [1]{#1}%
\providecommand \bibfnamefont [1]{#1}%
\providecommand \citenamefont [1]{#1}%
\providecommand \href@noop [0]{\@secondoftwo}%
\providecommand \href [0]{\begingroup \@sanitize@url \@href}%
\providecommand \@href[1]{\@@startlink{#1}\@@href}%
\providecommand \@@href[1]{\endgroup#1\@@endlink}%
\providecommand \@sanitize@url [0]{\catcode `\\12\catcode `\$12\catcode
  `\&12\catcode `\#12\catcode `\^12\catcode `\_12\catcode `\%12\relax}%
\providecommand \@@startlink[1]{}%
\providecommand \@@endlink[0]{}%
\providecommand \url  [0]{\begingroup\@sanitize@url \@url }%
\providecommand \@url [1]{\endgroup\@href {#1}{\urlprefix }}%
\providecommand \urlprefix  [0]{URL }%
\providecommand \Eprint [0]{\href }%
\providecommand \doibase [0]{http://dx.doi.org/}%
\providecommand \selectlanguage [0]{\@gobble}%
\providecommand \bibinfo  [0]{\@secondoftwo}%
\providecommand \bibfield  [0]{\@secondoftwo}%
\providecommand \translation [1]{[#1]}%
\providecommand \BibitemOpen [0]{}%
\providecommand \bibitemStop [0]{}%
\providecommand \bibitemNoStop [0]{.\EOS\space}%
\providecommand \EOS [0]{\spacefactor3000\relax}%
\providecommand \BibitemShut  [1]{\csname bibitem#1\endcsname}%
\let\auto@bib@innerbib\@empty
\bibitem [{\citenamefont {Stobi\ifmmode~\acute{n}\else \'{n}\fi{}ska}\ \emph
  {et~al.}(2008)\citenamefont {Stobi\ifmmode~\acute{n}\else \'{n}\fi{}ska},
  \citenamefont {Milburn},\ and\ \citenamefont
  {W\'odkiewicz}}]{StobMilbWodk08}%
  \BibitemOpen
  \bibfield  {author} {\bibinfo {author} {\bibfnamefont {M.}~\bibnamefont
  {Stobi\ifmmode~\acute{n}\else \'{n}\fi{}ska}}, \bibinfo {author}
  {\bibfnamefont {G.~J.}\ \bibnamefont {Milburn}}, \ and\ \bibinfo {author}
  {\bibfnamefont {K.}~\bibnamefont {W\'odkiewicz}},\ }\bibfield  {title}
  {\enquote {\bibinfo {title} {Wigner function evolution of quantum states in
  the presence of {self-Kerr} interaction},}\ }\href {\doibase
  10.1103/PhysRevA.78.013810} {\bibfield  {journal} {\bibinfo  {journal} {Phys.
  Rev. A}\ }\textbf {\bibinfo {volume} {78}},\ \bibinfo {pages} {013810}
  (\bibinfo {year} {2008})}\BibitemShut {NoStop}%
\bibitem [{\citenamefont {Genoni}\ \emph {et~al.}(2009)\citenamefont {Genoni},
  \citenamefont {Invernizzi},\ and\ \citenamefont {Paris}}]{GenoInvePari09}%
  \BibitemOpen
  \bibfield  {author} {\bibinfo {author} {\bibfnamefont {M.~G.}\ \bibnamefont
  {Genoni}}, \bibinfo {author} {\bibfnamefont {C.}~\bibnamefont {Invernizzi}},
  \ and\ \bibinfo {author} {\bibfnamefont {M.~G.~A.}\ \bibnamefont {Paris}},\
  }\bibfield  {title} {\enquote {\bibinfo {title} {Enhancement of parameter
  estimation by {Kerr} interaction},}\ }\href {\doibase
  10.1103/PhysRevA.80.033842} {\bibfield  {journal} {\bibinfo  {journal} {Phys.
  Rev. A}\ }\textbf {\bibinfo {volume} {80}},\ \bibinfo {pages} {033842}
  (\bibinfo {year} {2009})}\BibitemShut {NoStop}%
\bibitem [{\citenamefont {Rossi}\ \emph {et~al.}(2016)\citenamefont {Rossi},
  \citenamefont {Albarelli},\ and\ \citenamefont {Paris}}]{RossAlbaPari16}%
  \BibitemOpen
  \bibfield  {author} {\bibinfo {author} {\bibfnamefont {M.~A.~C.}\
  \bibnamefont {Rossi}}, \bibinfo {author} {\bibfnamefont {F.}~\bibnamefont
  {Albarelli}}, \ and\ \bibinfo {author} {\bibfnamefont {M.~G.~A.}\
  \bibnamefont {Paris}},\ }\bibfield  {title} {\enquote {\bibinfo {title}
  {Enhanced estimation of loss in the presence of {Kerr} nonlinearity},}\
  }\href {https://dx.doi.org/10.1103/PhysRevA.93.053805} {\bibfield  {journal}
  {\bibinfo  {journal} {Phys. Rev. A}\ }\textbf {\bibinfo {volume} {93}},\
  \bibinfo {pages} {053805} (\bibinfo {year} {2016})}\BibitemShut {NoStop}%
\bibitem [{\citenamefont {Roy}(2010)}]{Roy10}%
  \BibitemOpen
  \bibfield  {author} {\bibinfo {author} {\bibfnamefont {D.}~\bibnamefont
  {Roy}},\ }\bibfield  {title} {\enquote {\bibinfo {title} {Few-photon optical
  diode},}\ }\href {https://dx.doi.org/10.1103/PhysRevB.81.155117} {\bibfield
  {journal} {\bibinfo  {journal} {Phys. Rev. B}\ }\textbf {\bibinfo {volume}
  {81}},\ \bibinfo {pages} {155117} (\bibinfo {year} {2010})}\BibitemShut
  {NoStop}%
\bibitem [{\citenamefont {Mabuchi}(2012)}]{Mabu12}%
  \BibitemOpen
  \bibfield  {author} {\bibinfo {author} {\bibfnamefont {H.}~\bibnamefont
  {Mabuchi}},\ }\bibfield  {title} {\enquote {\bibinfo {title} {Qubit limit of
  cavity nonlinear optics},}\ }\href
  {https://dx.doi.org/10.1103/PhysRevA.85.015806} {\bibfield  {journal}
  {\bibinfo  {journal} {Phys. Rev. A}\ }\textbf {\bibinfo {volume} {85}},\
  \bibinfo {pages} {015806} (\bibinfo {year} {2012})}\BibitemShut {NoStop}%
\bibitem [{\citenamefont {Pachos}\ and\ \citenamefont
  {Chountasis}(2000)}]{PachChou00}%
  \BibitemOpen
  \bibfield  {author} {\bibinfo {author} {\bibfnamefont {J.}~\bibnamefont
  {Pachos}}\ and\ \bibinfo {author} {\bibfnamefont {S.}~\bibnamefont
  {Chountasis}},\ }\bibfield  {title} {\enquote {\bibinfo {title} {Optical
  holonomic quantum computer},}\ }\href
  {https://dx.doi.org/10.1103/PhysRevA.62.052318} {\bibfield  {journal}
  {\bibinfo  {journal} {Phys. Rev. A}\ }\textbf {\bibinfo {volume} {62}},\
  \bibinfo {pages} {052318} (\bibinfo {year} {2000})}\BibitemShut {NoStop}%
\bibitem [{\citenamefont {Knill}\ \emph {et~al.}(2001)\citenamefont {Knill},
  \citenamefont {Laflamme},\ and\ \citenamefont {Milburn}}]{KnilLaflMilb01}%
  \BibitemOpen
  \bibfield  {author} {\bibinfo {author} {\bibfnamefont {E.}~\bibnamefont
  {Knill}}, \bibinfo {author} {\bibfnamefont {R.}~\bibnamefont {Laflamme}}, \
  and\ \bibinfo {author} {\bibfnamefont {G.~J.}\ \bibnamefont {Milburn}},\
  }\bibfield  {title} {\enquote {\bibinfo {title} {A scheme for efficient
  quantum computation with linear optics},}\ }\href
  {http://dx.doi.org/10.1038/35051009} {\bibfield  {journal} {\bibinfo
  {journal} {Nature}\ }\textbf {\bibinfo {volume} {409}},\ \bibinfo {pages}
  {46} (\bibinfo {year} {2001})}\BibitemShut {NoStop}%
\bibitem [{\citenamefont {Nysteen}\ \emph {et~al.}(2017)\citenamefont
  {Nysteen}, \citenamefont {McCutcheon}, \citenamefont {Heuck}, \citenamefont
  {M\o{}rk},\ and\ \citenamefont {Englund}}]{NystMcCuHeuc17}%
  \BibitemOpen
  \bibfield  {author} {\bibinfo {author} {\bibfnamefont {A.}~\bibnamefont
  {Nysteen}}, \bibinfo {author} {\bibfnamefont {D.~P.~S.}\ \bibnamefont
  {McCutcheon}}, \bibinfo {author} {\bibfnamefont {M.}~\bibnamefont {Heuck}},
  \bibinfo {author} {\bibfnamefont {J.}~\bibnamefont {M\o{}rk}}, \ and\
  \bibinfo {author} {\bibfnamefont {D.~R.}\ \bibnamefont {Englund}},\
  }\bibfield  {title} {\enquote {\bibinfo {title} {Limitations of two-level
  emitters as nonlinearities in two-photon controlled-phase gates},}\ }\href
  {https://dx.doi.org/10.1103/PhysRevA.95.062304} {\bibfield  {journal}
  {\bibinfo  {journal} {Phys. Rev. A}\ }\textbf {\bibinfo {volume} {95}},\
  \bibinfo {pages} {062304} (\bibinfo {year} {2017})}\BibitemShut {NoStop}%
\bibitem [{\citenamefont {Drummond}\ and\ \citenamefont
  {Carter}(1987)}]{DrumCart87}%
  \BibitemOpen
  \bibfield  {author} {\bibinfo {author} {\bibfnamefont {P.~D.}\ \bibnamefont
  {Drummond}}\ and\ \bibinfo {author} {\bibfnamefont {S.~J.}\ \bibnamefont
  {Carter}},\ }\bibfield  {title} {\enquote {\bibinfo {title} {Quantum-field
  theory of squeezing in solitons},}\ }\href
  {http://dx.doi.org/10.1364/JOSAB.4.001565} {\bibfield  {journal} {\bibinfo
  {journal} {Journal of the Optical Society of America B}\ }\textbf {\bibinfo
  {volume} {4}},\ \bibinfo {pages} {1565} (\bibinfo {year} {1987})}\BibitemShut
  {NoStop}%
\bibitem [{\citenamefont {Lai}\ and\ \citenamefont {Haus}(1989)}]{LaiHaus89}%
  \BibitemOpen
  \bibfield  {author} {\bibinfo {author} {\bibfnamefont {Y.}~\bibnamefont
  {Lai}}\ and\ \bibinfo {author} {\bibfnamefont {H.~A.}\ \bibnamefont {Haus}},\
  }\bibfield  {title} {\enquote {\bibinfo {title} {Quantum theory of solitons
  in optical fibers. {II}. exact solution},}\ }\href {\doibase
  10.1103/PhysRevA.40.854} {\bibfield  {journal} {\bibinfo  {journal} {Phys.
  Rev. A}\ }\textbf {\bibinfo {volume} {40}},\ \bibinfo {pages} {854} (\bibinfo
  {year} {1989})}\BibitemShut {NoStop}%
\bibitem [{\citenamefont {Wright}(1990)}]{Wrig90}%
  \BibitemOpen
  \bibfield  {author} {\bibinfo {author} {\bibfnamefont {E.~M.}\ \bibnamefont
  {Wright}},\ }\bibfield  {title} {\enquote {\bibinfo {title} {Quantum theory
  of self-phase modulation},}\ }\bibfield  {booktitle} {\emph {\bibinfo
  {booktitle} {Journal of the Optical Society of America B}},\ }\href
  {http://dx.doi.org/10.1364/JOSAB.7.001142} {\bibfield  {journal} {\bibinfo
  {journal} {J. Opt. Soc. Am. B}\ }\textbf {\bibinfo {volume} {7}},\ \bibinfo
  {pages} {1142} (\bibinfo {year} {1990})}\BibitemShut {NoStop}%
\bibitem [{\citenamefont {Chiao}\ \emph {et~al.}(1991)\citenamefont {Chiao},
  \citenamefont {Deutsch},\ and\ \citenamefont {Garrison}}]{ChiaDeutGarr91}%
  \BibitemOpen
  \bibfield  {author} {\bibinfo {author} {\bibfnamefont {R.~Y.}\ \bibnamefont
  {Chiao}}, \bibinfo {author} {\bibfnamefont {I.~H.}\ \bibnamefont {Deutsch}},
  \ and\ \bibinfo {author} {\bibfnamefont {J.~C.}\ \bibnamefont {Garrison}},\
  }\bibfield  {title} {\enquote {\bibinfo {title} {Two-photon bound state in
  self-focusing media},}\ }\href
  {https://dx.doi.org/10.1103/PhysRevLett.67.1399} {\bibfield  {journal}
  {\bibinfo  {journal} {Phys. Rev. Lett.}\ }\textbf {\bibinfo {volume} {67}},\
  \bibinfo {pages} {1399} (\bibinfo {year} {1991})}\BibitemShut {NoStop}%
\bibitem [{\citenamefont {Blow}\ \emph {et~al.}(1991)\citenamefont {Blow},
  \citenamefont {Loudon},\ and\ \citenamefont {Phoenix}}]{BlowLoudPhoe91}%
  \BibitemOpen
  \bibfield  {author} {\bibinfo {author} {\bibfnamefont {K.~J.}\ \bibnamefont
  {Blow}}, \bibinfo {author} {\bibfnamefont {R.}~\bibnamefont {Loudon}}, \ and\
  \bibinfo {author} {\bibfnamefont {S.~J.~D.}\ \bibnamefont {Phoenix}},\
  }\bibfield  {title} {\enquote {\bibinfo {title} {Exact solution for quantum
  self-phase modulation},}\ }\bibfield  {booktitle} {\emph {\bibinfo
  {booktitle} {Journal of the Optical Society of America B}},\ }\href
  {http://dx.doi.org/10.1364/JOSAB.8.001750} {\bibfield  {journal} {\bibinfo
  {journal} {J. Opt. Soc. Am. B}\ }\textbf {\bibinfo {volume} {8}},\ \bibinfo
  {pages} {1750} (\bibinfo {year} {1991})}\BibitemShut {NoStop}%
\bibitem [{\citenamefont {Deutsch}\ \emph {et~al.}(1993)\citenamefont
  {Deutsch}, \citenamefont {Chiao},\ and\ \citenamefont
  {Garrison}}]{DeutChiaGarr93}%
  \BibitemOpen
  \bibfield  {author} {\bibinfo {author} {\bibfnamefont {I.~H.}\ \bibnamefont
  {Deutsch}}, \bibinfo {author} {\bibfnamefont {R.~Y.}\ \bibnamefont {Chiao}},
  \ and\ \bibinfo {author} {\bibfnamefont {J.~C.}\ \bibnamefont {Garrison}},\
  }\bibfield  {title} {\enquote {\bibinfo {title} {Two-photon bound states: The
  diphoton bullet in dispersive self-focusing media},}\ }\href
  {http://dx.doi.org/10.1103/PhysRevA.47.3330} {\bibfield  {journal} {\bibinfo
  {journal} {Phys. Rev. A}\ }\textbf {\bibinfo {volume} {47}},\ \bibinfo
  {pages} {3330} (\bibinfo {year} {1993})}\BibitemShut {NoStop}%
\bibitem [{\citenamefont {Joneckis}\ and\ \citenamefont
  {Shapiro}(1993)}]{JoneShap93}%
  \BibitemOpen
  \bibfield  {author} {\bibinfo {author} {\bibfnamefont {L.~G.}\ \bibnamefont
  {Joneckis}}\ and\ \bibinfo {author} {\bibfnamefont {J.~H.}\ \bibnamefont
  {Shapiro}},\ }\bibfield  {title} {\enquote {\bibinfo {title} {Quantum
  propagation in a {Kerr} medium: lossless, dispersionless fiber},}\ }\href
  {http://dx.doi.org/10.1364/JOSAB.10.001102} {\bibfield  {journal} {\bibinfo
  {journal} {J. Opt. Soc. Am. B}\ }\textbf {\bibinfo {volume} {10}},\ \bibinfo
  {pages} {1102} (\bibinfo {year} {1993})}\BibitemShut {NoStop}%
\bibitem [{\citenamefont {Abram}\ and\ \citenamefont
  {Cohen}(1994)}]{AbraCohe94}%
  \BibitemOpen
  \bibfield  {author} {\bibinfo {author} {\bibfnamefont {I.}~\bibnamefont
  {Abram}}\ and\ \bibinfo {author} {\bibfnamefont {E.}~\bibnamefont {Cohen}},\
  }\bibfield  {title} {\enquote {\bibinfo {title} {Quantum propagation of light
  in a {Kerr} medium renormalization},}\ }\href
  {http://dx.doi.org/10.1080/09500349414550851} {\bibfield  {journal} {\bibinfo
   {journal} {Journal of Modern Optics}\ }\textbf {\bibinfo {volume} {41}},\
  \bibinfo {pages} {847} (\bibinfo {year} {1994})}\BibitemShut {NoStop}%
\bibitem [{\citenamefont {Boivin}\ \emph {et~al.}(1994)\citenamefont {Boivin},
  \citenamefont {K\"artner},\ and\ \citenamefont {Haus}}]{BoivKartHaus94}%
  \BibitemOpen
  \bibfield  {author} {\bibinfo {author} {\bibfnamefont {L.}~\bibnamefont
  {Boivin}}, \bibinfo {author} {\bibfnamefont {F.~X.}\ \bibnamefont
  {K\"artner}}, \ and\ \bibinfo {author} {\bibfnamefont {H.~A.}\ \bibnamefont
  {Haus}},\ }\bibfield  {title} {\enquote {\bibinfo {title} {Analytical
  solution to the quantum field theory of self-phase modulation with a finite
  response time},}\ }\href {https://dx.doi.org/10.1103/PhysRevLett.73.240}
  {\bibfield  {journal} {\bibinfo  {journal} {Phys. Rev. Lett.}\ }\textbf
  {\bibinfo {volume} {73}},\ \bibinfo {pages} {240} (\bibinfo {year}
  {1994})}\BibitemShut {NoStop}%
\bibitem [{\citenamefont {Wei\ss{}l}\ \emph {et~al.}(2015)\citenamefont
  {Wei\ss{}l}, \citenamefont {K\"ung}, \citenamefont {Dumur}, \citenamefont
  {Feofanov}, \citenamefont {Matei}, \citenamefont {Naud}, \citenamefont
  {Buisson}, \citenamefont {Hekking},\ and\ \citenamefont
  {Guichard}}]{WeisKungDumu15}%
  \BibitemOpen
  \bibfield  {author} {\bibinfo {author} {\bibfnamefont {T.}~\bibnamefont
  {Wei\ss{}l}}, \bibinfo {author} {\bibfnamefont {B.}~\bibnamefont {K\"ung}},
  \bibinfo {author} {\bibfnamefont {E.}~\bibnamefont {Dumur}}, \bibinfo
  {author} {\bibfnamefont {A.~K.}\ \bibnamefont {Feofanov}}, \bibinfo {author}
  {\bibfnamefont {I.}~\bibnamefont {Matei}}, \bibinfo {author} {\bibfnamefont
  {C.}~\bibnamefont {Naud}}, \bibinfo {author} {\bibfnamefont {O.}~\bibnamefont
  {Buisson}}, \bibinfo {author} {\bibfnamefont {F.~W.~J.}\ \bibnamefont
  {Hekking}}, \ and\ \bibinfo {author} {\bibfnamefont {W.}~\bibnamefont
  {Guichard}},\ }\bibfield  {title} {\enquote {\bibinfo {title} {Kerr
  coefficients of plasma resonances in {Josephson} junction chains},}\ }\href
  {https://dx.doi.org/10.1103/PhysRevB.92.104508} {\bibfield  {journal}
  {\bibinfo  {journal} {Phys. Rev. B}\ }\textbf {\bibinfo {volume} {92}},\
  \bibinfo {pages} {104508} (\bibinfo {year} {2015})}\BibitemShut {NoStop}%
\bibitem [{\citenamefont {Parkins}(2016)}]{Park16}%
  \BibitemOpen
  \bibfield  {author} {\bibinfo {author} {\bibfnamefont {S.}~\bibnamefont
  {Parkins}},\ }\bibfield  {title} {\enquote {\bibinfo {title} {Optical quantum
  logic at the ultimate limit},}\ }\href
  {https://physics.aps.org/articles/v9/129} {\bibfield  {journal} {\bibinfo
  {journal} {Physics}\ }\textbf {\bibinfo {volume} {9}},\ \bibinfo {pages}
  {129} (\bibinfo {year} {2016})}\BibitemShut {NoStop}%
\bibitem [{\citenamefont {Gorshkov}\ \emph {et~al.}(2011)\citenamefont
  {Gorshkov}, \citenamefont {Otterbach}, \citenamefont {Fleischhauer},
  \citenamefont {Pohl},\ and\ \citenamefont {Lukin}}]{GorsOtteFlei11}%
  \BibitemOpen
  \bibfield  {author} {\bibinfo {author} {\bibfnamefont {A.~V.}\ \bibnamefont
  {Gorshkov}}, \bibinfo {author} {\bibfnamefont {J.}~\bibnamefont {Otterbach}},
  \bibinfo {author} {\bibfnamefont {M.}~\bibnamefont {Fleischhauer}}, \bibinfo
  {author} {\bibfnamefont {T.}~\bibnamefont {Pohl}}, \ and\ \bibinfo {author}
  {\bibfnamefont {M.~D.}\ \bibnamefont {Lukin}},\ }\bibfield  {title} {\enquote
  {\bibinfo {title} {Photon-photon interactions via {R}ydberg blockade},}\
  }\href {https://dx.doi.org/10.1103/PhysRevLett.107.133602} {\bibfield
  {journal} {\bibinfo  {journal} {Phys. Rev. Lett.}\ }\textbf {\bibinfo
  {volume} {107}},\ \bibinfo {pages} {133602} (\bibinfo {year}
  {2011})}\BibitemShut {NoStop}%
\bibitem [{\citenamefont {Bienias}\ and\ \citenamefont
  {B{\"u}chler}(2016)}]{BienBuch16}%
  \BibitemOpen
  \bibfield  {author} {\bibinfo {author} {\bibfnamefont {P.}~\bibnamefont
  {Bienias}}\ and\ \bibinfo {author} {\bibfnamefont {H.~P.}\ \bibnamefont
  {B{\"u}chler}},\ }\bibfield  {title} {\enquote {\bibinfo {title} {Quantum
  theory of {Kerr} nonlinearity with {Rydberg} slow light polaritons},}\ }\href
  {https://doi.org/10.1088/1367-2630/aa50c3} {\bibfield  {journal} {\bibinfo
  {journal} {New Journal of Physics}\ }\textbf {\bibinfo {volume} {18}},\
  \bibinfo {pages} {123026} (\bibinfo {year} {2016})}\BibitemShut {NoStop}%
\bibitem [{\citenamefont {Lahad}\ and\ \citenamefont
  {Firstenberg}(2017)}]{LahaFirs17}%
  \BibitemOpen
  \bibfield  {author} {\bibinfo {author} {\bibfnamefont {O.}~\bibnamefont
  {Lahad}}\ and\ \bibinfo {author} {\bibfnamefont {O.}~\bibnamefont
  {Firstenberg}},\ }\bibfield  {title} {\enquote {\bibinfo {title} {Induced
  cavities for photonic quantum gates},}\ }\href
  {https://dx.doi.org/10.1103/PhysRevLett.119.113601} {\bibfield  {journal}
  {\bibinfo  {journal} {Phys. Rev. Lett.}\ }\textbf {\bibinfo {volume} {119}},\
  \bibinfo {pages} {113601} (\bibinfo {year} {2017})}\BibitemShut {NoStop}%
\bibitem [{\citenamefont {Lee}\ \emph {et~al.}(2015)\citenamefont {Lee},
  \citenamefont {Noh}, \citenamefont {Schetakis},\ and\ \citenamefont
  {Angelakis}}]{LeeNohSche15}%
  \BibitemOpen
  \bibfield  {author} {\bibinfo {author} {\bibfnamefont {C.}~\bibnamefont
  {Lee}}, \bibinfo {author} {\bibfnamefont {C.}~\bibnamefont {Noh}}, \bibinfo
  {author} {\bibfnamefont {N.}~\bibnamefont {Schetakis}}, \ and\ \bibinfo
  {author} {\bibfnamefont {D.~G.}\ \bibnamefont {Angelakis}},\ }\bibfield
  {title} {\enquote {\bibinfo {title} {Few-photon transport in many-body
  photonic systems: A scattering approach},}\ }\href
  {http://dx.doi.org/10.1103/PhysRevA.92.063817} {\bibfield  {journal}
  {\bibinfo  {journal} {Phys. Rev. A}\ }\textbf {\bibinfo {volume} {92}},\
  \bibinfo {pages} {063817} (\bibinfo {year} {2015})}\BibitemShut {NoStop}%
\bibitem [{\citenamefont {See}\ \emph {et~al.}(2017)\citenamefont {See},
  \citenamefont {Noh},\ and\ \citenamefont {Angelakis}}]{SeeNohAnge17}%
  \BibitemOpen
  \bibfield  {author} {\bibinfo {author} {\bibfnamefont {T.~F.}\ \bibnamefont
  {See}}, \bibinfo {author} {\bibfnamefont {C.}~\bibnamefont {Noh}}, \ and\
  \bibinfo {author} {\bibfnamefont {D.~G.}\ \bibnamefont {Angelakis}},\
  }\bibfield  {title} {\enquote {\bibinfo {title} {Diagrammatic approach to
  multiphoton scattering},}\ }\href
  {https://dx.doi.org/10.1103/PhysRevA.95.053845} {\bibfield  {journal}
  {\bibinfo  {journal} {Phys. Rev. A}\ }\textbf {\bibinfo {volume} {95}},\
  \bibinfo {pages} {053845} (\bibinfo {year} {2017})}\BibitemShut {NoStop}%
\bibitem [{\citenamefont {Pedersen}\ and\ \citenamefont
  {Pletyukhov}(2017)}]{PedePlet17}%
  \BibitemOpen
  \bibfield  {author} {\bibinfo {author} {\bibfnamefont {K.~G.~L.}\
  \bibnamefont {Pedersen}}\ and\ \bibinfo {author} {\bibfnamefont
  {M.}~\bibnamefont {Pletyukhov}},\ }\bibfield  {title} {\enquote {\bibinfo
  {title} {Few-photon scattering on {Bose-Hubbard} lattices},}\ }\href
  {https://dx.doi.org/10.1103/PhysRevA.96.023815} {\bibfield  {journal}
  {\bibinfo  {journal} {Phys. Rev. A}\ }\textbf {\bibinfo {volume} {96}},\
  \bibinfo {pages} {023815} (\bibinfo {year} {2017})}\BibitemShut {NoStop}%
\bibitem [{\citenamefont {Waks}\ and\ \citenamefont
  {Vuckovic}(2006)}]{WaksVuck06}%
  \BibitemOpen
  \bibfield  {author} {\bibinfo {author} {\bibfnamefont {E.}~\bibnamefont
  {Waks}}\ and\ \bibinfo {author} {\bibfnamefont {J.}~\bibnamefont
  {Vuckovic}},\ }\bibfield  {title} {\enquote {\bibinfo {title} {Dispersive
  properties and large {K}err nonlinearities using dipole-induced transparency
  in a single-sided cavity},}\ }\href {\doibase 10.1103/PhysRevA.73.041803}
  {\bibfield  {journal} {\bibinfo  {journal} {Phys. Rev. A}\ }\textbf {\bibinfo
  {volume} {73}},\ \bibinfo {pages} {041803} (\bibinfo {year}
  {2006})}\BibitemShut {NoStop}%
\bibitem [{\citenamefont {Koshino}(2008)}]{Kosh08}%
  \BibitemOpen
  \bibfield  {author} {\bibinfo {author} {\bibfnamefont {K.}~\bibnamefont
  {Koshino}},\ }\bibfield  {title} {\enquote {\bibinfo {title} {Multiphoton
  wave function after {Kerr} interaction},}\ }\href
  {https://dx.doi.org/10.1103/PhysRevA.78.023820} {\bibfield  {journal}
  {\bibinfo  {journal} {Phys. Rev. A}\ }\textbf {\bibinfo {volume} {78}},\
  \bibinfo {pages} {023820} (\bibinfo {year} {2008})}\BibitemShut {NoStop}%
\bibitem [{\citenamefont {Liao}\ and\ \citenamefont {Law}(2010)}]{LiaoLaw10}%
  \BibitemOpen
  \bibfield  {author} {\bibinfo {author} {\bibfnamefont {J.-Q.}\ \bibnamefont
  {Liao}}\ and\ \bibinfo {author} {\bibfnamefont {C.~K.}\ \bibnamefont {Law}},\
  }\bibfield  {title} {\enquote {\bibinfo {title} {Correlated two-photon
  transport in a one-dimensional waveguide side-coupled to a nonlinear
  cavity},}\ }\href {\doibase 10.1103/PhysRevA.82.053836} {\bibfield  {journal}
  {\bibinfo  {journal} {Phys. Rev. A}\ }\textbf {\bibinfo {volume} {82}},\
  \bibinfo {pages} {053836} (\bibinfo {year} {2010})}\BibitemShut {NoStop}%
\bibitem [{\citenamefont {Xu}\ \emph {et~al.}(2013)\citenamefont {Xu},
  \citenamefont {Rephaeli},\ and\ \citenamefont {Fan}}]{XuRephFan13}%
  \BibitemOpen
  \bibfield  {author} {\bibinfo {author} {\bibfnamefont {S.}~\bibnamefont
  {Xu}}, \bibinfo {author} {\bibfnamefont {E.}~\bibnamefont {Rephaeli}}, \ and\
  \bibinfo {author} {\bibfnamefont {S.}~\bibnamefont {Fan}},\ }\bibfield
  {title} {\enquote {\bibinfo {title} {Analytic properties of two-photon
  scattering matrix in integrated quantum systems determined by the cluster
  decomposition principle},}\ }\href {\doibase 10.1103/PhysRevLett.111.223602}
  {\bibfield  {journal} {\bibinfo  {journal} {Phys. Rev. Lett.}\ }\textbf
  {\bibinfo {volume} {111}},\ \bibinfo {pages} {223602} (\bibinfo {year}
  {2013})}\BibitemShut {NoStop}%
\bibitem [{\citenamefont {Shapiro}(2006)}]{Shap06}%
  \BibitemOpen
  \bibfield  {author} {\bibinfo {author} {\bibfnamefont {J.~H.}\ \bibnamefont
  {Shapiro}},\ }\bibfield  {title} {\enquote {\bibinfo {title} {Single-photon
  {K}err nonlinearities do not help quantum computation},}\ }\href
  {http://dx.doi.org//10.1103/PhysRevA.73.062305} {\bibfield  {journal}
  {\bibinfo  {journal} {Phys. Rev. A}\ }\textbf {\bibinfo {volume} {73}},\
  \bibinfo {pages} {062305} (\bibinfo {year} {2006})}\BibitemShut {NoStop}%
\bibitem [{\citenamefont {Gea-Banacloche}(2010)}]{GeaBana10}%
  \BibitemOpen
  \bibfield  {author} {\bibinfo {author} {\bibfnamefont {J.}~\bibnamefont
  {Gea-Banacloche}},\ }\bibfield  {title} {\enquote {\bibinfo {title}
  {Impossibility of large phase shifts via the giant {K}err effect with
  single-photon wave packets},}\ }\href {\doibase 10.1103/PhysRevA.81.043823}
  {\bibfield  {journal} {\bibinfo  {journal} {Phys. Rev. A}\ }\textbf {\bibinfo
  {volume} {81}},\ \bibinfo {pages} {043823} (\bibinfo {year}
  {2010})}\BibitemShut {NoStop}%
\bibitem [{\citenamefont {Xu}\ and\ \citenamefont {Fan}(2017)}]{XuFan17}%
  \BibitemOpen
  \bibfield  {author} {\bibinfo {author} {\bibfnamefont {S.}~\bibnamefont
  {Xu}}\ and\ \bibinfo {author} {\bibfnamefont {S.}~\bibnamefont {Fan}},\
  }\bibfield  {title} {\enquote {\bibinfo {title} {Generalized cluster
  decomposition principle illustrated in waveguide quantum electrodynamics},}\
  }\href {https://link.aps.org/doi/10.1103/PhysRevA.95.063809} {\bibfield
  {journal} {\bibinfo  {journal} {Phys. Rev. A}\ }\textbf {\bibinfo {volume}
  {95}},\ \bibinfo {pages} {063809} (\bibinfo {year} {2017})}\BibitemShut
  {NoStop}%
\bibitem [{\citenamefont {S{\'a}nchez-Burillo}\ \emph
  {et~al.}(2018)\citenamefont {S{\'a}nchez-Burillo}, \citenamefont {Cadarso},
  \citenamefont {Mart{\'\i}n-Moreno}, \citenamefont {Garc{\'\i}a-Ripoll},\ and\
  \citenamefont {Zueco}}]{SancCadaMart18}%
  \BibitemOpen
  \bibfield  {author} {\bibinfo {author} {\bibfnamefont {E.}~\bibnamefont
  {S{\'a}nchez-Burillo}}, \bibinfo {author} {\bibfnamefont {A.}~\bibnamefont
  {Cadarso}}, \bibinfo {author} {\bibfnamefont {L.}~\bibnamefont
  {Mart{\'\i}n-Moreno}}, \bibinfo {author} {\bibfnamefont {J.~J.}\ \bibnamefont
  {Garc{\'\i}a-Ripoll}}, \ and\ \bibinfo {author} {\bibfnamefont
  {D.}~\bibnamefont {Zueco}},\ }\bibfield  {title} {\enquote {\bibinfo {title}
  {Emergent causality and the n -photon scattering matrix in waveguide qed},}\
  }\href {https://doi.org/10.1088/1367-2630/aa9cc2} {\bibfield  {journal}
  {\bibinfo  {journal} {New Journal of Physics}\ }\textbf {\bibinfo {volume}
  {20}},\ \bibinfo {pages} {013017} (\bibinfo {year} {2018})}\BibitemShut
  {NoStop}%
\bibitem [{\citenamefont {Brod}\ \emph {et~al.}(2016)\citenamefont {Brod},
  \citenamefont {Combes},\ and\ \citenamefont {Gea-Banacloche}}]{BrodComb16a}%
  \BibitemOpen
  \bibfield  {author} {\bibinfo {author} {\bibfnamefont {D.~J.}\ \bibnamefont
  {Brod}}, \bibinfo {author} {\bibfnamefont {J.}~\bibnamefont {Combes}}, \ and\
  \bibinfo {author} {\bibfnamefont {J.}~\bibnamefont {Gea-Banacloche}},\
  }\bibfield  {title} {\enquote {\bibinfo {title} {Two photons co- and
  counter-propagating through {$N$} cross-{K}err sites},}\ }\href
  {http://dx.doi.org/10.1103/PhysRevA.94.023833} {\bibfield  {journal}
  {\bibinfo  {journal} {Phys. Rev. A}\ }\textbf {\bibinfo {volume} {94}},\
  \bibinfo {pages} {023833} (\bibinfo {year} {2016})}\BibitemShut {NoStop}%
\bibitem [{\citenamefont {Brod}\ and\ \citenamefont
  {Combes}(2016)}]{BrodComb16b}%
  \BibitemOpen
  \bibfield  {author} {\bibinfo {author} {\bibfnamefont {D.~J.}\ \bibnamefont
  {Brod}}\ and\ \bibinfo {author} {\bibfnamefont {J.}~\bibnamefont {Combes}},\
  }\bibfield  {title} {\enquote {\bibinfo {title} {Passive {CPHASE} gate via
  {cross-{K}err} nonlinearities},}\ }\href
  {http://dx.doi.org/10.1103/PhysRevLett.117.080502} {\bibfield  {journal}
  {\bibinfo  {journal} {Phys. Rev. Lett.}\ }\textbf {\bibinfo {volume} {117}},\
  \bibinfo {pages} {080502} (\bibinfo {year} {2016})}\BibitemShut {NoStop}%
\bibitem [{\citenamefont {Gopalan}\ \emph {et~al.}(1994)\citenamefont
  {Gopalan}, \citenamefont {Rice},\ and\ \citenamefont
  {Sigrist}}]{GopaRiceSigr94}%
  \BibitemOpen
  \bibfield  {author} {\bibinfo {author} {\bibfnamefont {S.}~\bibnamefont
  {Gopalan}}, \bibinfo {author} {\bibfnamefont {T.~M.}\ \bibnamefont {Rice}}, \
  and\ \bibinfo {author} {\bibfnamefont {M.}~\bibnamefont {Sigrist}},\
  }\bibfield  {title} {\enquote {\bibinfo {title} {Spin ladders with spin gaps:
  A description of a class of cuprates},}\ }\href {\doibase
  https://dx.doi.org/10.1103/PhysRevB.49.8901} {\bibfield  {journal} {\bibinfo
  {journal} {Phys. Rev. B}\ }\textbf {\bibinfo {volume} {49}},\ \bibinfo
  {pages} {8901} (\bibinfo {year} {1994})}\BibitemShut {NoStop}%
\bibitem [{\citenamefont {S{\'o}lyom}(2000)}]{Soly00}%
  \BibitemOpen
  \bibfield  {author} {\bibinfo {author} {\bibfnamefont {J.}~\bibnamefont
  {S{\'o}lyom}},\ }\bibfield  {title} {\enquote {\bibinfo {title} {Critical and
  massive phases in spin chains and spin ladders},}\ }\href {\doibase
  https://dx.doi.org/10.12693/APhysPolA.97.81} {\bibfield  {journal} {\bibinfo
  {journal} {Acta Physica Polonica A}\ }\textbf {\bibinfo {volume} {97}},\
  \bibinfo {pages} {81} (\bibinfo {year} {2000})}\BibitemShut {NoStop}%
\bibitem [{\citenamefont {Mila}(2000)}]{Mila00}%
  \BibitemOpen
  \bibfield  {author} {\bibinfo {author} {\bibfnamefont {F.}~\bibnamefont
  {Mila}},\ }\bibfield  {title} {\enquote {\bibinfo {title} {Quantum spin
  liquids},}\ }\href {https://doi.org/10.1088/0143-0807/21/6/302} {\bibfield
  {journal} {\bibinfo  {journal} {European Journal of Physics}\ }\textbf
  {\bibinfo {volume} {21}},\ \bibinfo {pages} {499} (\bibinfo {year}
  {2000})}\BibitemShut {NoStop}%
\bibitem [{\citenamefont {Zhou}\ \emph {et~al.}(2017)\citenamefont {Zhou},
  \citenamefont {Kanoda},\ and\ \citenamefont {Ng}}]{ZhouKanoNg17}%
  \BibitemOpen
  \bibfield  {author} {\bibinfo {author} {\bibfnamefont {Y.}~\bibnamefont
  {Zhou}}, \bibinfo {author} {\bibfnamefont {K.}~\bibnamefont {Kanoda}}, \ and\
  \bibinfo {author} {\bibfnamefont {T.-K.}\ \bibnamefont {Ng}},\ }\bibfield
  {title} {\enquote {\bibinfo {title} {Quantum spin liquid states},}\ }\href
  {\doibase https://dx.doi.org/10.1103/RevModPhys.89.025003} {\bibfield
  {journal} {\bibinfo  {journal} {Rev. Mod. Phys.}\ }\textbf {\bibinfo {volume}
  {89}},\ \bibinfo {pages} {025003} (\bibinfo {year} {2017})}\BibitemShut
  {NoStop}%
\bibitem [{\citenamefont {Collett}\ and\ \citenamefont
  {Gardiner}(1984)}]{CollGard84}%
  \BibitemOpen
  \bibfield  {author} {\bibinfo {author} {\bibfnamefont {M.~J.}\ \bibnamefont
  {Collett}}\ and\ \bibinfo {author} {\bibfnamefont {C.~W.}\ \bibnamefont
  {Gardiner}},\ }\bibfield  {title} {\enquote {\bibinfo {title} {Squeezing of
  intracavity and traveling-wave light fields produced in parametric
  amplification},}\ }\href {\doibase 10.1103/PhysRevA.30.1386} {\bibfield
  {journal} {\bibinfo  {journal} {Phys. Rev. A}\ }\textbf {\bibinfo {volume}
  {30}},\ \bibinfo {pages} {1386} (\bibinfo {year} {1984})}\BibitemShut
  {NoStop}%
\bibitem [{\citenamefont {Gardiner}\ and\ \citenamefont
  {Collett}(1985)}]{GardColl85}%
  \BibitemOpen
  \bibfield  {author} {\bibinfo {author} {\bibfnamefont {C.~W.}\ \bibnamefont
  {Gardiner}}\ and\ \bibinfo {author} {\bibfnamefont {M.~J.}\ \bibnamefont
  {Collett}},\ }\bibfield  {title} {\enquote {\bibinfo {title} {Input and
  output in damped quantum systems: Quantum stochastic differential equations
  and the master equation},}\ }\href {\doibase 10.1103/PhysRevA.31.3761}
  {\bibfield  {journal} {\bibinfo  {journal} {Phys. Rev. A}\ }\textbf {\bibinfo
  {volume} {31}},\ \bibinfo {pages} {3761} (\bibinfo {year}
  {1985})}\BibitemShut {NoStop}%
\bibitem [{\citenamefont {Gough}\ and\ \citenamefont
  {James}(2009{\natexlab{a}})}]{GougJame09}%
  \BibitemOpen
  \bibfield  {author} {\bibinfo {author} {\bibfnamefont {J.}~\bibnamefont
  {Gough}}\ and\ \bibinfo {author} {\bibfnamefont {M.}~\bibnamefont {James}},\
  }\bibfield  {title} {\enquote {\bibinfo {title} {The series product and its
  application to quantum feedforward and feedback networks},}\ }\href
  {http://dx.doi.org/10.1109/TAC.2009.2031205} {\bibfield  {journal} {\bibinfo
  {journal} {Automatic Control, IEEE Transactions on}\ }\textbf {\bibinfo
  {volume} {54}},\ \bibinfo {pages} {2530} (\bibinfo {year}
  {2009}{\natexlab{a}})}\BibitemShut {NoStop}%
\bibitem [{\citenamefont {Gough}\ and\ \citenamefont
  {James}(2009{\natexlab{b}})}]{GougJame09a}%
  \BibitemOpen
  \bibfield  {author} {\bibinfo {author} {\bibfnamefont {J.}~\bibnamefont
  {Gough}}\ and\ \bibinfo {author} {\bibfnamefont {M.~R.}\ \bibnamefont
  {James}},\ }\bibfield  {title} {\enquote {\bibinfo {title} {{Quantum feedback
  networks: {Hamiltonian} formulation}},}\ }\href
  {http://dx.doi.org/10.1007/s00220-008-0698-8} {\bibfield  {journal} {\bibinfo
   {journal} {Comm. Math. Phys.}\ }\textbf {\bibinfo {volume} {287}},\ \bibinfo
  {pages} {1109} (\bibinfo {year} {2009}{\natexlab{b}})}\BibitemShut {NoStop}%
\bibitem [{\citenamefont {Carmichael}(1993)}]{Carm93}%
  \BibitemOpen
  \bibfield  {author} {\bibinfo {author} {\bibfnamefont {H.~J.}\ \bibnamefont
  {Carmichael}},\ }\bibfield  {title} {\enquote {\bibinfo {title} {Quantum
  trajectory theory for cascaded open systems},}\ }\href
  {http://dx.doi.org/10.1103/PhysRevLett.70.2273} {\bibfield  {journal}
  {\bibinfo  {journal} {Phys. Rev. Lett.}\ }\textbf {\bibinfo {volume} {70}},\
  \bibinfo {pages} {2273} (\bibinfo {year} {1993})}\BibitemShut {NoStop}%
\bibitem [{\citenamefont {Gardiner}(1993)}]{Gard93}%
  \BibitemOpen
  \bibfield  {author} {\bibinfo {author} {\bibfnamefont {C.~W.}\ \bibnamefont
  {Gardiner}},\ }\bibfield  {title} {\enquote {\bibinfo {title} {Driving a
  quantum system with the output field from another driven quantum system},}\
  }\href {\doibase 10.1103/PhysRevLett.70.2269} {\bibfield  {journal} {\bibinfo
   {journal} {Phys. Rev. Lett.}\ }\textbf {\bibinfo {volume} {70}},\ \bibinfo
  {pages} {2269} (\bibinfo {year} {1993})}\BibitemShut {NoStop}%
\bibitem [{\citenamefont {Combes}\ \emph {et~al.}(2017)\citenamefont {Combes},
  \citenamefont {Kerckhoff},\ and\ \citenamefont {Sarovar}}]{CombKercSaro17}%
  \BibitemOpen
  \bibfield  {author} {\bibinfo {author} {\bibfnamefont {J.}~\bibnamefont
  {Combes}}, \bibinfo {author} {\bibfnamefont {J.}~\bibnamefont {Kerckhoff}}, \
  and\ \bibinfo {author} {\bibfnamefont {M.}~\bibnamefont {Sarovar}},\
  }\bibfield  {title} {\enquote {\bibinfo {title} {The {SLH} framework for
  modeling quantum input-output networks},}\ }\href
  {http://dx.doi.org/10.1080/23746149.2017.1343097} {\bibfield  {journal}
  {\bibinfo  {journal} {Advances in Physics: X}\ }\textbf {\bibinfo {volume}
  {2}},\ \bibinfo {pages} {784} (\bibinfo {year} {2017})}\BibitemShut {NoStop}%
\bibitem [{\citenamefont {Dalton}\ \emph {et~al.}(1999)\citenamefont {Dalton},
  \citenamefont {Barnett},\ and\ \citenamefont {Knight}}]{DaltBarnKnig99}%
  \BibitemOpen
  \bibfield  {author} {\bibinfo {author} {\bibfnamefont {B.~J.}\ \bibnamefont
  {Dalton}}, \bibinfo {author} {\bibfnamefont {S.~M.}\ \bibnamefont {Barnett}},
  \ and\ \bibinfo {author} {\bibfnamefont {P.~L.}\ \bibnamefont {Knight}},\
  }\bibfield  {title} {\enquote {\bibinfo {title} {A quantum scattering theory
  approach to quantum-optical measurements},}\ }\href
  {https://dx.doi.org/10.1080/09500349908230404} {\bibfield  {journal}
  {\bibinfo  {journal} {Journal of Modern Optics}\ }\textbf {\bibinfo {volume}
  {46}},\ \bibinfo {pages} {1107} (\bibinfo {year} {1999})}\BibitemShut
  {NoStop}%
\bibitem [{\citenamefont {Fan}\ \emph {et~al.}(2010)\citenamefont {Fan},
  \citenamefont {Kocaba\c{s}},\ and\ \citenamefont {Shen}}]{FanKocaShen10}%
  \BibitemOpen
  \bibfield  {author} {\bibinfo {author} {\bibfnamefont {S.}~\bibnamefont
  {Fan}}, \bibinfo {author} {\bibfnamefont {{\c{S}}.~E.}\ \bibnamefont
  {Kocaba\c{s}}}, \ and\ \bibinfo {author} {\bibfnamefont {J.-T.}\ \bibnamefont
  {Shen}},\ }\bibfield  {title} {\enquote {\bibinfo {title} {Input-output
  formalism for few-photon transport in one-dimensional nanophotonic waveguides
  coupled to a qubit},}\ }\href {\doibase 10.1103/PhysRevA.82.063821}
  {\bibfield  {journal} {\bibinfo  {journal} {Phys. Rev. A}\ }\textbf {\bibinfo
  {volume} {82}},\ \bibinfo {pages} {063821} (\bibinfo {year}
  {2010})}\BibitemShut {NoStop}%
\bibitem [{\citenamefont {Pletyukhov}\ and\ \citenamefont
  {Gritsev}(2012)}]{PletGrit12}%
  \BibitemOpen
  \bibfield  {author} {\bibinfo {author} {\bibfnamefont {M.}~\bibnamefont
  {Pletyukhov}}\ and\ \bibinfo {author} {\bibfnamefont {V.}~\bibnamefont
  {Gritsev}},\ }\bibfield  {title} {\enquote {\bibinfo {title} {Scattering of
  massless particles in one-dimensional chiral channel},}\ }\href
  {https://doi.org/10.1088/1367-2630/14/9/095028} {\bibfield  {journal}
  {\bibinfo  {journal} {New Journal of Physics}\ }\textbf {\bibinfo {volume}
  {14}},\ \bibinfo {pages} {095028} (\bibinfo {year} {2012})}\BibitemShut
  {NoStop}%
\bibitem [{\citenamefont {Roy}\ \emph {et~al.}(2017)\citenamefont {Roy},
  \citenamefont {Wilson},\ and\ \citenamefont {Firstenberg}}]{RoyWilsFirs16}%
  \BibitemOpen
  \bibfield  {author} {\bibinfo {author} {\bibfnamefont {D.}~\bibnamefont
  {Roy}}, \bibinfo {author} {\bibfnamefont {C.~M.}\ \bibnamefont {Wilson}}, \
  and\ \bibinfo {author} {\bibfnamefont {O.}~\bibnamefont {Firstenberg}},\
  }\bibfield  {title} {\enquote {\bibinfo {title} {Colloquium: Strongly
  interacting photons in one-dimensional continuum},}\ }\href
  {https://dx.doi.org/10.1103/RevModPhys.89.021001} {\bibfield  {journal}
  {\bibinfo  {journal} {Rev. Mod. Phys.}\ }\textbf {\bibinfo {volume} {89}},\
  \bibinfo {pages} {021001} (\bibinfo {year} {2017})}\BibitemShut {NoStop}%
\bibitem [{\citenamefont {Ma\ifmmode~\check{s}\else \v{s}\fi{}alas}\ and\
  \citenamefont {Fleischhauer}(2004)}]{MassFlei04}%
  \BibitemOpen
  \bibfield  {author} {\bibinfo {author} {\bibfnamefont {M.}~\bibnamefont
  {Ma\ifmmode~\check{s}\else \v{s}\fi{}alas}}\ and\ \bibinfo {author}
  {\bibfnamefont {M.}~\bibnamefont {Fleischhauer}},\ }\bibfield  {title}
  {\enquote {\bibinfo {title} {Scattering of dark-state polaritons in optical
  lattices and quantum phase gate for photons},}\ }\href
  {http://dx.doi.org/10.1103/PhysRevA.69.061801} {\bibfield  {journal}
  {\bibinfo  {journal} {Phys. Rev. A}\ }\textbf {\bibinfo {volume} {69}},\
  \bibinfo {pages} {061801} (\bibinfo {year} {2004})}\BibitemShut {NoStop}%
\bibitem [{\citenamefont {Milburn}(1989)}]{Milb89}%
  \BibitemOpen
  \bibfield  {author} {\bibinfo {author} {\bibfnamefont {G.~J.}\ \bibnamefont
  {Milburn}},\ }\bibfield  {title} {\enquote {\bibinfo {title} {Quantum optical
  {F}redkin gate},}\ }\href {\doibase 10.1103/PhysRevLett.62.2124} {\bibfield
  {journal} {\bibinfo  {journal} {Phys. Rev. Lett.}\ }\textbf {\bibinfo
  {volume} {62}},\ \bibinfo {pages} {2124} (\bibinfo {year}
  {1989})}\BibitemShut {NoStop}%
\bibitem [{\citenamefont {Chuang}\ and\ \citenamefont
  {Yamamoto}(1995)}]{ChuaYama95}%
  \BibitemOpen
  \bibfield  {author} {\bibinfo {author} {\bibfnamefont {I.~L.}\ \bibnamefont
  {Chuang}}\ and\ \bibinfo {author} {\bibfnamefont {Y.}~\bibnamefont
  {Yamamoto}},\ }\bibfield  {title} {\enquote {\bibinfo {title} {Simple quantum
  computer},}\ }\href {\doibase 10.1103/PhysRevA.52.3489} {\bibfield  {journal}
  {\bibinfo  {journal} {Phys. Rev. A}\ }\textbf {\bibinfo {volume} {52}},\
  \bibinfo {pages} {3489} (\bibinfo {year} {1995})}\BibitemShut {NoStop}%
\bibitem [{\citenamefont {Chudzicki}\ \emph {et~al.}(2013)\citenamefont
  {Chudzicki}, \citenamefont {Chuang},\ and\ \citenamefont
  {Shapiro}}]{ChudChuaShap13}%
  \BibitemOpen
  \bibfield  {author} {\bibinfo {author} {\bibfnamefont {C.}~\bibnamefont
  {Chudzicki}}, \bibinfo {author} {\bibfnamefont {I.~L.}\ \bibnamefont
  {Chuang}}, \ and\ \bibinfo {author} {\bibfnamefont {J.~H.}\ \bibnamefont
  {Shapiro}},\ }\bibfield  {title} {\enquote {\bibinfo {title} {Deterministic
  and cascadable conditional phase gate for photonic qubits},}\ }\href
  {\doibase 10.1103/PhysRevA.87.042325} {\bibfield  {journal} {\bibinfo
  {journal} {Phys. Rev. A}\ }\textbf {\bibinfo {volume} {87}},\ \bibinfo
  {pages} {042325} (\bibinfo {year} {2013})}\BibitemShut {NoStop}%
\bibitem [{\citenamefont {Thompson}\ \emph {et~al.}(2016)\citenamefont
  {Thompson}, \citenamefont {Gokler}, \citenamefont {Lloyd},\ and\
  \citenamefont {Shor}}]{ThomGoklLloy15}%
  \BibitemOpen
  \bibfield  {author} {\bibinfo {author} {\bibfnamefont {K.~F.}\ \bibnamefont
  {Thompson}}, \bibinfo {author} {\bibfnamefont {C.}~\bibnamefont {Gokler}},
  \bibinfo {author} {\bibfnamefont {S.}~\bibnamefont {Lloyd}}, \ and\ \bibinfo
  {author} {\bibfnamefont {P.~W.}\ \bibnamefont {Shor}},\ }\bibfield  {title}
  {\enquote {\bibinfo {title} {Time independent universal computing with spin
  chains: quantum plinko machine},}\ }\href
  {https://dx.doi.org/10.1088/1367-2630/18/7/073044} {\bibfield  {journal}
  {\bibinfo  {journal} {New Journal of Physics}\ }\textbf {\bibinfo {volume}
  {18}},\ \bibinfo {pages} {073044} (\bibinfo {year} {2016})}\BibitemShut
  {NoStop}%
\bibitem [{\citenamefont {Gorshkov}\ \emph {et~al.}(2010)\citenamefont
  {Gorshkov}, \citenamefont {Otterbach}, \citenamefont {Demler}, \citenamefont
  {Fleischhauer},\ and\ \citenamefont {Lukin}}]{GorsOtteDeml10}%
  \BibitemOpen
  \bibfield  {author} {\bibinfo {author} {\bibfnamefont {A.~V.}\ \bibnamefont
  {Gorshkov}}, \bibinfo {author} {\bibfnamefont {J.}~\bibnamefont {Otterbach}},
  \bibinfo {author} {\bibfnamefont {E.}~\bibnamefont {Demler}}, \bibinfo
  {author} {\bibfnamefont {M.}~\bibnamefont {Fleischhauer}}, \ and\ \bibinfo
  {author} {\bibfnamefont {M.~D.}\ \bibnamefont {Lukin}},\ }\bibfield  {title}
  {\enquote {\bibinfo {title} {Photonic phase gate via an exchange of fermionic
  spin waves in a spin chain},}\ }\href
  {https://dx.doi.org/10.1103/PhysRevLett.105.060502} {\bibfield  {journal}
  {\bibinfo  {journal} {Phys. Rev. Lett.}\ }\textbf {\bibinfo {volume} {105}},\
  \bibinfo {pages} {060502} (\bibinfo {year} {2010})}\BibitemShut {NoStop}%
\bibitem [{\citenamefont {Datta}(1997)}]{Datt97}%
  \BibitemOpen
  \bibfield  {author} {\bibinfo {author} {\bibfnamefont {S.}~\bibnamefont
  {Datta}},\ }\href@noop {} {\emph {\bibinfo {title} {Electronic transport in
  mesoscopic systems}}}\ (\bibinfo  {publisher} {Cambridge university press},\
  \bibinfo {year} {1997})\BibitemShut {NoStop}%
\bibitem [{\citenamefont {Dutra}(2005)}]{Dutr05}%
  \BibitemOpen
  \bibfield  {author} {\bibinfo {author} {\bibfnamefont {S.~M.}\ \bibnamefont
  {Dutra}},\ }\href@noop {} {\emph {\bibinfo {title} {Cavity Quantum
  Electrodynamics: The Strange Theory of Light in a Box}}}\ (\bibinfo
  {publisher} {John Wiley \& Sons},\ \bibinfo {year} {2005})\ p.\ \bibinfo
  {pages} {408}\BibitemShut {NoStop}%
\bibitem [{\citenamefont {Cohen}\ and\ \citenamefont
  {M{\o}lmer}(2018)}]{CoheMolm}%
  \BibitemOpen
  \bibfield  {author} {\bibinfo {author} {\bibfnamefont {I.}~\bibnamefont
  {Cohen}}\ and\ \bibinfo {author} {\bibfnamefont {K.}~\bibnamefont
  {M{\o}lmer}},\ }\bibfield  {title} {\enquote {\bibinfo {title} {Deterministic
  quantum network for distributed entanglement and quantum computation},}\
  }\href {http://dx.doi.org/10.1103/PhysRevA.98.030302} {\bibfield  {journal}
  {\bibinfo  {journal} {Phys. Rev. A}\ }\textbf {\bibinfo {volume} {98}},\
  \bibinfo {pages} {030302} (\bibinfo {year} {2018})}\BibitemShut {NoStop}%
\bibitem [{\citenamefont {Rudolph}(2017)}]{Rudolph17}%
  \BibitemOpen
  \bibfield  {author} {\bibinfo {author} {\bibfnamefont {T.}~\bibnamefont
  {Rudolph}},\ }\bibfield  {title} {\enquote {\bibinfo {title} {Why {I} am
  optimistic about the silicon-photonic route to quantum computing},}\ }\href
  {\doibase 10.1063/1.4976737} {\bibfield  {journal} {\bibinfo  {journal} {APL
  Photonics}\ }\textbf {\bibinfo {volume} {2}},\ \bibinfo {pages} {030901}
  (\bibinfo {year} {2017})}\BibitemShut {NoStop}%
\bibitem [{\citenamefont {Childs}\ \emph {et~al.}(2005)\citenamefont {Childs},
  \citenamefont {Leung},\ and\ \citenamefont {Nielsen}}]{ChilLeunNiel05}%
  \BibitemOpen
  \bibfield  {author} {\bibinfo {author} {\bibfnamefont {A.~M.}\ \bibnamefont
  {Childs}}, \bibinfo {author} {\bibfnamefont {D.~W.}\ \bibnamefont {Leung}}, \
  and\ \bibinfo {author} {\bibfnamefont {M.~A.}\ \bibnamefont {Nielsen}},\
  }\bibfield  {title} {\enquote {\bibinfo {title} {Unified derivations of
  measurement-based schemes for quantum computation},}\ }\href
  {https://dx.doi.org/10.1103/PhysRevA.71.032318} {\bibfield  {journal}
  {\bibinfo  {journal} {Phys. Rev. A}\ }\textbf {\bibinfo {volume} {71}},\
  \bibinfo {pages} {032318} (\bibinfo {year} {2005})}\BibitemShut {NoStop}%
\bibitem [{\citenamefont {Gimeno-Segovia}\ \emph {et~al.}(2015)\citenamefont
  {Gimeno-Segovia}, \citenamefont {Shadbolt}, \citenamefont {Browne},\ and\
  \citenamefont {Rudolph}}]{Gimeno-Segovia15}%
  \BibitemOpen
  \bibfield  {author} {\bibinfo {author} {\bibfnamefont {M.}~\bibnamefont
  {Gimeno-Segovia}}, \bibinfo {author} {\bibfnamefont {P.}~\bibnamefont
  {Shadbolt}}, \bibinfo {author} {\bibfnamefont {D.~E.}\ \bibnamefont
  {Browne}}, \ and\ \bibinfo {author} {\bibfnamefont {T.}~\bibnamefont
  {Rudolph}},\ }\bibfield  {title} {\enquote {\bibinfo {title} {From
  three-photon {Greenberger-Horne-Zeilinger} states to ballistic universal
  quantum computation},}\ }\href {\doibase 10.1103/PhysRevLett.115.020502}
  {\bibfield  {journal} {\bibinfo  {journal} {Phys. Rev. Lett.}\ }\textbf
  {\bibinfo {volume} {115}},\ \bibinfo {pages} {020502} (\bibinfo {year}
  {2015})}\BibitemShut {NoStop}%
\bibitem [{\citenamefont {Pandey}\ \emph {et~al.}(2013)\citenamefont {Pandey},
  \citenamefont {Jiang}, \citenamefont {Combes},\ and\ \citenamefont
  {Caves}}]{PandJianComb13}%
  \BibitemOpen
  \bibfield  {author} {\bibinfo {author} {\bibfnamefont {S.}~\bibnamefont
  {Pandey}}, \bibinfo {author} {\bibfnamefont {Z.}~\bibnamefont {Jiang}},
  \bibinfo {author} {\bibfnamefont {J.}~\bibnamefont {Combes}}, \ and\ \bibinfo
  {author} {\bibfnamefont {C.~M.}\ \bibnamefont {Caves}},\ }\bibfield  {title}
  {\enquote {\bibinfo {title} {Quantum limits on probabilistic amplifiers},}\
  }\href {https://dx.doi.org/10.1103/PhysRevA.88.033852} {\bibfield  {journal}
  {\bibinfo  {journal} {Phys. Rev. A}\ }\textbf {\bibinfo {volume} {88}},\
  \bibinfo {pages} {033852} (\bibinfo {year} {2013})}\BibitemShut {NoStop}%
\bibitem [{\citenamefont {Uskov}\ \emph {et~al.}(2009)\citenamefont {Uskov},
  \citenamefont {Kaplan}, \citenamefont {Smith}, \citenamefont {Huver},\ and\
  \citenamefont {Dowling}}]{Uskov09}%
  \BibitemOpen
  \bibfield  {author} {\bibinfo {author} {\bibfnamefont {D.~B.}\ \bibnamefont
  {Uskov}}, \bibinfo {author} {\bibfnamefont {L.}~\bibnamefont {Kaplan}},
  \bibinfo {author} {\bibfnamefont {A.~M.}\ \bibnamefont {Smith}}, \bibinfo
  {author} {\bibfnamefont {S.~D.}\ \bibnamefont {Huver}}, \ and\ \bibinfo
  {author} {\bibfnamefont {J.~P.}\ \bibnamefont {Dowling}},\ }\bibfield
  {title} {\enquote {\bibinfo {title} {Maximal success probabilities of
  linear-optical quantum gates},}\ }\href {\doibase 10.1103/PhysRevA.79.042326}
  {\bibfield  {journal} {\bibinfo  {journal} {Phys. Rev. A}\ }\textbf {\bibinfo
  {volume} {79}},\ \bibinfo {pages} {042326} (\bibinfo {year}
  {2009})}\BibitemShut {NoStop}%
\bibitem [{\citenamefont {Lodahl}\ \emph {et~al.}(2017)\citenamefont {Lodahl},
  \citenamefont {Mahmoodian}, \citenamefont {Stobbe}, \citenamefont
  {Rauschenbeutel}, \citenamefont {Schneeweiss}, \citenamefont {Volz},
  \citenamefont {Pichler},\ and\ \citenamefont {Zoller}}]{Lodahl2017}%
  \BibitemOpen
  \bibfield  {author} {\bibinfo {author} {\bibfnamefont {P.}~\bibnamefont
  {Lodahl}}, \bibinfo {author} {\bibfnamefont {S.}~\bibnamefont {Mahmoodian}},
  \bibinfo {author} {\bibfnamefont {S.}~\bibnamefont {Stobbe}}, \bibinfo
  {author} {\bibfnamefont {A.}~\bibnamefont {Rauschenbeutel}}, \bibinfo
  {author} {\bibfnamefont {P.}~\bibnamefont {Schneeweiss}}, \bibinfo {author}
  {\bibfnamefont {J.}~\bibnamefont {Volz}}, \bibinfo {author} {\bibfnamefont
  {H.}~\bibnamefont {Pichler}}, \ and\ \bibinfo {author} {\bibfnamefont
  {P.}~\bibnamefont {Zoller}},\ }\bibfield  {title} {\enquote {\bibinfo {title}
  {Chiral quantum optics},}\ }\href {https://doi.org/10.1038/nature21037}
  {\bibfield  {journal} {\bibinfo  {journal} {Nature}\ }\textbf {\bibinfo
  {volume} {541}},\ \bibinfo {pages} {473} (\bibinfo {year}
  {2017})}\BibitemShut {NoStop}%
\bibitem [{\citenamefont {Konyk}\ and\ \citenamefont
  {Gea-Banacloche}(2018)}]{KonyGeaB18}%
  \BibitemOpen
  \bibfield  {author} {\bibinfo {author} {\bibfnamefont {W.}~\bibnamefont
  {Konyk}}\ and\ \bibinfo {author} {\bibfnamefont {J.}~\bibnamefont
  {Gea-Banacloche}},\ }\bibfield  {title} {\enquote {\bibinfo {title} {Passive,
  deterministic photonic cphase gate via two-level systems},}\ }\href
  {https://arxiv.org/abs/1807.07224} {\bibfield  {journal} {\bibinfo  {journal}
  {arXiv:1807.07224}\ } (\bibinfo {year} {2018})}\BibitemShut {NoStop}%
\end{thebibliography}%
\vspace{5em}


\appendix

\section{Computing fidelities}\label{app:fidelities}

For completeness, here we describe how to obtain the average gate fidelities from the S-matrices in \cref{eq:2pInfsres,eq:SM1pNsres}. For a more in-depth discussion, see \cite{BrodComb16a,BrodComb16b}

The first step is to choose an input wave packet shape. Here we assume input photons to have Gaussian profiles, given by 
\begin{equation}
\xi_{\textrm{in}}(\omega) = \frac{1}{(2\pi \sigma^2)^{1/4}}\exp\left[ -\frac{(\omega -\omega_c)^2}{4 \sigma^2} \right],
\end{equation}
where $\omega_0$ is the detuning (i.e.\ the carrier frequency is $\omega_c=\Delta+\omega_0$) and $\sigma$ is the bandwidth. These can be mapped to the corresponding output wavepackets using the S-matrix:
\begin{align*}
\xi_{\textrm{out}}^{(1)}(\nu) & = \int d\omega S_{\omega,\nu} \xi_{\textrm{in}}(\omega), \\
\xi_{\textrm{out}}^{(2)}(\nu_1,\nu_2) & = \int d \omega_1 d\omega_2 S_{\omega_1,\omega_2,\nu_1,\nu_2} \xi_{\textrm{in}}(\omega_1) \xi_{\textrm{in}}(\omega_2).
\end{align*}

We remark that all integrations must be performed only in half the frequency space, e.g.\ for $\omega_1 \geq \omega_2$, since bosonic statistics means that $\ket{\omega_1^+ \omega_2^+}$ and $\ket{\omega_2^+ \omega_1^+}$ are actually the same state. The average gate fidelity can now be computed as
\begin{align} \label{eq:fidelity0b}
F(\phi) := \int d\psi \bra{\psi} S_{\rm id}(\phi)\dg S_{\rm act}\op{\psi}{\psi}  S_{\rm act}\dg S_{\rm id}(\phi) \ket{\psi}
\end{align}
where the integration is over the two-qubit Haar measure, and $S_{\rm id}(\phi)$ and $S_{\rm act}$ are the desired S-matrix and the one that describes our system, respectively. Following the arguments in \cite{BrodComb16b}, we choose to ignore the single-photon deformation, as there are error-correction techniques that allow us to circumvent it. Concretely, this means our ideal S-matrix $S_{\rm id}(\phi)$ is of the form (\ref{eq:idealSmatrix}), but with $C_{\nu_1,\nu_2} = e^{i\phi} \barG{\nu_1}^{2N} \barG{\nu_2}^{2N}$ rather than simply $C_{\nu_1,\nu_2} = e^{i\phi}$.

By using the fact that we ignore single-photon deformation and that our operation conserves photon number, we can show that \cref{eq:fidelity0b} reduces simply to 
\begin{equation} \label{eq:fidelity3}
F(\phi) = \frac{1}{10} \left( 6 + 3 \textrm{Re}(e^{i \phi} \mathcal{F}) + |\mathcal{F}|^2 \right),
\end{equation}
where $\mathcal{F}$ is the overlap between the output single- and two-photon output wave packets:
\begin{equation}
\mathcal{F} = \int \conj{[\xi_{\rm out}^{(1)}(\nu_1)\xi_{\rm out}^{(1)}(\nu_2)]} \xi_{\rm out}^{(2)}(\nu_1,\nu_2) d\nu_1 d\nu_2.
\end{equation}
From this expression we performed numerical integrations for different numbers of sites and photon bandwidths, and obtained the fidelities reported in the main text. By computing fidelities for $N = 4 \ldots 20$ we were able to fit the expected asymptotic behavior of these quantities reported in \cref{eq:asympt}.

\section{Comparison with probabilistic linear-optical gates}

In this Appendix we discuss some practical aspects of our proposal and compare it with probabilistic linear-optical gates, such as the KLM scheme. We begin by pointing out two issues that complicate such a comparison. First, non-linear and linear-optical logic gates use different resources, and which one is more efficient depends heavily on underlying available technologies. Second, both approaches still require much theoretical improvement, e.g.\ especially-tailored quantum error-correction. Thus, we caution the reader not to read too much into the following comparisons. We intend to carry out a full analysis of the experimental feasability of our approach in future work. For now, we direct the reader to \cite{BrodComb16b,BrodComb16a} for open questions regarding our proposal, and to \cite{Rudolph17} for a (somewhat optimisc) review of the state-of-the-art of linear-optical quantum computing. 

Consider the following three practical aspects of quantum-optical computing: 
\begin{enumerate}
\item Losses;
\item Break-even point for deterministic gates vs.\ probabilistic gates (such as in the KLM scheme \cite{KnilLaflMilb01});
\item Overhead in the number of optical elements.
\end{enumerate}

{\em Losses.} Let us assume that at each scattering site there are photon losses $\eta$, which may arise from propagation losses or emission into nonguided modes. We model such losses as a beam splitter into an unobserved mode:
\begin{align}
{\rm BS} = \left(\begin{array}{cc}\sqrt{1-\eta^2} & -\eta \\  \eta & \sqrt{1-\eta^2} \end{array}\right).
\end{align}
For simplicity, we make two assumptions: first, that losses commute through into a single beam splitter at the end of the chain, and second that losses are sufficiently small such that $ N \eta\ll1$. Thus, after $N$ cross-Kerr sites we have losses corresponding to a final beamsplitter given by $({\rm BS})^{2N}$, leading to a loss probability of approximately $4N^2\eta^2$ per photon.

In the context of quantum computation there is another perspective on loss. From the threshold theorem, we know that it suffices for gates to be above a certain fixed fidelity, and so we can choose some {\em constant} $N$, say 5 or 12. Furthermore, with a measurement-based approach such as \cite{ChilLeunNiel05}, each photon crosses only 3 CPHASE gates in the entire computation. Thus, there is always $\eta$ small enough such that any photon survives with high probability regardless of the computation size. Thus, even if losses are problematic for near-term implementations, they are not a fundamental obstacle for our proposal and can be resolved with technological improvements. This feature is also shared by linear-optical proposals for a similar reason, as discussed in \cite{Rudolph17,Gimeno-Segovia15}.

{\em Deterministic break-even point.} We now make a simplistic comparison between probabilistic (i.e.\ KLM) and deterministic optical gates. Suppose we perform $D$ two-qubit gates in sequence. As figure of merit we choose the probability-fidelity $pF$ product \cite{PandJianComb13}, which one can argue represents the overall probability to achieve the desired goal. For deterministic gates, the success probability is $P_{\rm det} = 1$, while each gate has fidelity $F_{\rm det}=1-\epsilon$. For $D$ gates, $(pF)_{\rm det}=(1 - \epsilon)^D$. In the probabilistic case, gates have $P_{\rm prob} = 1/\delta $ with $\delta \geq 1$ and fidelity $F_{\rm prob}=1$. In the standard KLM scheme, $P_{\rm prob} = 1/16$ (at the cost of two ancilla photons). There is numerical evidence to suggest that the maximum achievable probability for unit-fidelity two-qubit gates is 2/27 \cite{Uskov09}, at least if one is restricted to a few ancilla photons. At any rate, the $pF$ product for probabilistic gates is $(pF)_{\rm prob} = 1/\delta^{D}$. Thus, for any $\epsilon<1 - \delta$ we have $(pF)_{\rm Det}>(pF)_{\rm non-Det}$. Since, as we have shown, we can achieve arbitrarily low $\epsilon$, this suggests that high-quality deterministic gates always outperform probabilistic gates.

This analysis leaves out a few important points. First, the most promising approach for linear-optical quantum computing is based on so-called fusion gates, which work with probability 3/4 \cite{Gimeno-Segovia15}. Their action (and failure modes) are not exactly the same as the KLM CPHASE gate, and it is not obvious how to include them in the above comparison. Second, the success of linear-optical gates are always heralded at \textit{each} gate. Thus, when failure happens you can abort the computation, or let standard error-correction techniques deal with it. Alternatively, recent approaches \cite{Gimeno-Segovia15} use probabilistic gates to construct random graph states, and use them for computation. This is a promising approach not amenable to the above analysis.

{\em Number of elements.} According to \cite{Rudolph17}, the most up-to-date estimate on the resource overhead of linear-optical quantum computing is the following: supposing one has access to a deterministic source of entangled three-photon states (which is an open challenge), a cluster-state architecture would induce an overhead of 20 physical photons per logical qubit, would not need quantum memories, and every photon would only ever go through a constant number of optical elements (limiting the detrimental effect of losses).

Our approach, in its current form, requires 5 (12) interaction sites to achieve a fidelity of $95\%$ ($99\%$), plus a similar number of circulators unless one uses tricks from ``chiral quantum optics'' \cite{Lodahl2017}. As discussed in the main text, if all interactions are polarization-independent this number can be halved. By using measurement-based quantum computing, our proposal also benefits from a constant-depth (hence low loss). Another benefit is that all elements are passive, whereas linear-optical proposals rely heavily on measurement adaptation and side classical processing (with the caveat that these elements would likely be necessary for quantum error correction or measurement-based quantum computation). Finally, we point out that linear optics has received two decades of theoretical work that pushed down its resource overhead by several orders of magnitude, whereas quantum computing based on the Kerr effect has not received the same amount of attention, partly due to previous work suggesting it was impossible. We believe that there is much theoretical work to be done to simplify our proposal and bring it to a feasibility level comparable with current technology, see e.g. the recent work of \citet{KonyGeaB18} which improves the resource count on our cross-Kerr proposal.

Ultimately, we leave for the reader and for experiments to decide which approach is more feasible, and how this comparison will evolve with future technological advancements. We echo the feeling expressed in \cite{Rudolph17} that, at the end of the day, a full-fledged quantum computer, with quantum channels between its components and connected to a quantum communications network, will likely require a hybrid scheme involving matter-based and optics-based implementations, motivating research and development in a diversity of directions.

\end{document}